\documentclass[preprint,12pt]{elsarticle}

\usepackage{bibunits}
\usepackage{amssymb}
\usepackage{amsmath}
\usepackage[version=4,arrows=pgf]{mhchem}
\usepackage[T1]{fontenc}
\usepackage{graphicx}
\usepackage[hidelinks]{hyperref}

\begin{document}


\begin{bibunit}[elsarticle-num]

\begin{frontmatter}


\title{Decoupling support-dependent transport profiles from molecular water motion}

\author[inst1]{Jannik Mehlis}
\author[inst1,inst2]{Matthias Wessling\corref{cor1}}

\ead{manuscripts.cvt@avt.rwth-aachen.de}
\cortext[cor1]{Corresponding author:}

\affiliation[inst1]{organization={Chemical Process Engineering AVT.CVT, RWTH Aachen University},
            addressline={Forckenbeckstraße 51}, 
            city={Aachen},
            postcode={52074}, 
            country={Germany}}

\affiliation[inst2]{organization={DWI - Leibniz Institute for Interactive Materials e.V},
            addressline={Forckenbeckstraße 50}, 
            city={Aachen},
            postcode={52074}, 
            country={Germany}}

\begin{abstract}

Pressure and concentration profiles obtained from non-equilibrium molecular dynamics (NEMD) simulations are commonly used to infer water transport mechanisms in dense and swollen polymer membranes. However, the mechanical restraints required to stabilize the simulated membrane can alter the resulting profiles and potentially affect their mechanistic interpretation. Here, we examine whether restraint-induced differences in pressure and water concentration profiles are accompanied by changes in molecular water dynamics. Crosslinked poly(ethylene glycol) diacrylate (PEGDA) membranes were simulated using three mechanical support strategies: a graphene support lattice, frozen membrane atoms localized near the permeate interface, and homogeneously distributed frozen membrane atoms. Although these restraint schemes produced markedly different water concentration and pressure profiles, ranging from solution--diffusion-like to pore-flow-like, the overall water flux remained comparable. More importantly, molecular water dynamics were largely insensitive to the restraint strategy. Interfacial water exchange was strongly bidirectional and substantially exceeded the net permeation flux, membrane water diffusivities were similar, and complete membrane crossings occurred in both directions. Local water associations were short-lived: initially neighboring molecules progressively separated, while directional velocity correlations decayed rapidly with both time and distance. These results demonstrate that substantially different pressure and concentration profiles can arise despite similar underlying molecular water dynamics. Pressure and concentration profiles alone are therefore insufficient to identify molecular transport mechanisms in mechanically restrained NEMD simulations. For the PEGDA system studied here, molecular water motion is dominated by stochastic diffusion with a small net directional bias.

\end{abstract}

%


\end{frontmatter}



\section{Introduction}

    Liquid pressure-driven membrane separation processes, including nanofiltration and reverse osmosis, rely on selective solvent transport through dense or swollen polymer membranes~\cite{yangReviewReverseOsmosis2019}. Despite their technological importance and maturity, the molecular origin of water permeation remains debated~\cite{heiranianMechanismsModelsWater2023, zhuMultiscaleModellingTransport2024}. In the solution--diffusion (SD) model, water partitions into the membrane according to its chemical potential and subsequently undergoes stochastic diffusion down a concentration gradient; desorption at the permeate interface is likewise governed by thermodynamic equilibrium~\cite{lonsdale1965transport, wijmansSolutiondiffusionModelReview1995, paul2004reformulation}. By contrast, pore-flow (PF) descriptions emphasize pressure-driven transport through connected pathways, with a hydraulic pressure gradient contributing to the driving force~\cite{bakerMembraneTechnologyApplications}. The difficulty is that macroscopic observables such as flux--pressure relations, water concentration profiles, and apparent pressure gradients do not directly reveal the molecular events responsible for transport. In addition, membrane-mechanical assumptions and experimental or computational boundary conditions can generate profiles that appear consistent with either SD- or PF-like transport~\cite{wijmansPermeationExperimentsReveal2026, fregerSolutiondiffusionModelRumors2024, hegdeTwophaseModelThat2022, fregerSolutiondiffusionmechanicsModelDerivation2025, spiegler1966thermodynamics}. A profile that agrees with one transport model therefore does not necessarily reveal how individual water molecules move through the membrane.
    
    Recent experimental studies have renewed interest in the molecular origin of solvent transport through dense polymer membranes by reporting pressure-dependent solvent concentration gradients~\cite{sujananiHydraulicPermeationinducedWater2024, ogiegloNHexaneInducedSwelling2013}, nonlinear permeation~\cite{reimundExperimentalObservationNonlinear2025a}, and pressure-induced compaction~\cite{fanIrreversibleCompactionGoverns2026}. Crosslinked poly(ethylene glycol) diacrylate (PEGDA) hydrogels provide a useful model system for examining this problem because their water uptake and permeation behavior have been characterized experimentally~\cite{reimundExperimentalObservationNonlinear2025a, juPreparationCharacterizationCrosslinked2009}, while ion sorption and diffusion have been quantified as functions of hydration~\cite{jangInfluenceWaterContent2020, jangInfluenceWaterContent2019}. Molecular simulations of PEGDA and related hydrogel networks have additionally resolved water and ion diffusion, polymer--ion interactions, crosslinking effects, and hydration heterogeneity~\cite{zofchakCationPolymerInteractions2023, rukmaniMolecularModelingComplex2019, wuEffectCrossLinkingDiffusion2009, jangMechanicalTransportProperties2007, luoEffectLoopDefects2020}. These experimental and simulation data provide independent structural and transport benchmarks for the atomistic model used here. Previous PEGDA simulations, however, have mainly considered equilibrium or fully periodic hydrogel systems rather than pressure-driven transport through a mechanically supported membrane slab.
    
    Non-equilibrium molecular dynamics (NEMD) simulations provide a direct route to resolving pressure-driven transport at molecular resolution. Previous NEMD studies of dense polymer membranes have reported nearly uniform water concentrations and approximately linear pressure profiles and interpreted these signatures as evidence for PF-like transport~\cite{wangNonequilibriumMolecularSimulations2025, heMolecularSimulationsElucidate2024a, liwangWaterTransportReverse2023}. More recent work has established, however, that the same spatially resolved observables are strongly affected by the mechanical constraints imposed on the membrane~\cite{fregerSolutiondiffusionmechanicsModelDerivation2025, marioniNonequilibriumSimulationsHydraulic2026a}. Restraining selected membrane atoms can introduce localized stresses and produce PF-like profiles, whereas support-lattice geometries distribute the mechanical load differently and can recover concentration gradients and approximately constant internal pressures that are compatible with SD theory~\cite{fregerSolutiondiffusionmechanicsModelDerivation2025, marioniNonequilibriumSimulationsHydraulic2026a}. This prior work shows that macroscopic NEMD profiles depend on the mechanical boundary conditions. It does not resolve whether the different profiles are accompanied by different molecular water dynamics, or whether similar molecular motion can produce them under different restraints.
    
    Here, we address this question for pressure-driven water permeation through crosslinked PEGDA membranes using three support strategies commonly employed in NEMD simulations: a graphene support lattice, frozen membrane atoms localized near the permeate interface, and homogeneously distributed frozen membrane atoms. The test is straightforward. If restraint-dependent pressure and concentration profiles represent genuinely different transport mechanisms, changing the support strategy should also change molecular measures of water transport. If the molecular dynamics remain similar despite different profiles, the profiles alone cannot establish the molecular transport mechanism. We therefore compare pressure and concentration profiles with interfacial water exchange, intramembrane mobility, individual molecular trajectories, the persistence of local water associations, and directional correlations. In this way, the known effect of mechanical support on the calculated profiles can be separated from its effect on molecular water motion.

\section{Methods} \label{sec:Methods}

    \begin{figure}[ht]
    \centering
    \includegraphics[width=\linewidth]{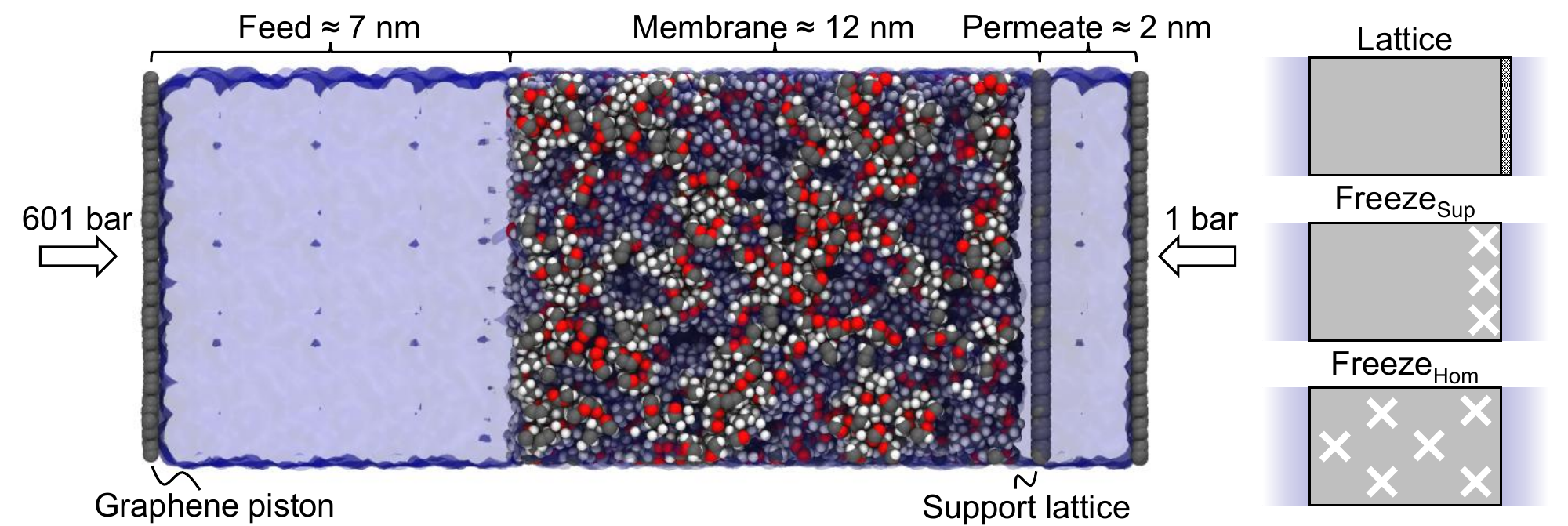}
    \caption{Simulation setup for the NEMD simulations. The central region contains the crosslinked PEGDA membrane, with the feed compartment on the left and the permeate compartment on the right. Pressure-driven transport was induced by applying external forces to two graphene pistons positioned at the outer boundaries of the simulation system. In the example shown, the feed and permeate pressures were $601\,\mathrm{bar}$ and $1\,\mathrm{bar}$, respectively, resulting in a pressure difference of $600\,\mathrm{bar}$ across the membrane. A graphene support lattice was placed on the permeate side to provide mechanical support. The three support strategies investigated are illustrated on the right: a graphene support lattice (Lattice), as shown in the full simulation setup on the left; frozen membrane atoms on the permeate interface ($\mathrm{Freeze}_{\mathrm{Sup}}$), indicated by white crosses; and homogeneously distributed frozen membrane atoms throughout the membrane ($\mathrm{Freeze}_{\mathrm{Hom}}$).}
    \label{fig:fig1}
    \end{figure}
    
    PEGDA membranes were generated using an automated crosslinking procedure and equilibrated to a water content of $37\,\mathrm{vol.-\%}$, consistent with experimentally characterized PEGDA systems~\cite{reimundExperimentalObservationNonlinear2025a, jangInfluenceWaterContent2020, jangInfluenceWaterContent2019}. The crosslinking workflow is available on \href{http://permalink.avt.rwth-aachen.de/?id=472286}{GitLab}. Five independently generated membrane configurations were simulated to quantify system-to-system variability. All simulations were performed using Gromacs 2024.4~\cite{abrahamGROMACSHighPerformance2015, pallTacklingExascaleSoftware2015} with parameters from the OPLS-AA force field~\cite{jorgensenDevelopmentTestingOPLS1996b}.

    Pressure-driven water permeation was investigated using NEMD simulations comprising explicit feed and permeate reservoirs and graphene pistons that imposed the pressure difference, as illustrated in Figure~\ref{fig:fig1}. To evaluate the influence of membrane support on water transport, three support strategies were compared at a fixed pressure difference of $600\,\mathrm{bar}$: a graphene support lattice on the permeate side (Lattice), membrane atoms frozen locally near the permeate interface ($\mathrm{Freeze}_{\mathrm{Sup}}$), and homogeneously distributed frozen membrane atoms ($\mathrm{Freeze}_{\mathrm{Hom}}$). These support strategies are illustrated in the right panel of Figure~\ref{fig:fig1}. The Lattice configuration was subsequently used to investigate the pressure dependence of water transport at pressure differences of $150$, $300$, $600$, and $900\,\mathrm{bar}$. These pressure differences substantially exceed those typically applied in experimental reverse osmosis but are commonly required in NEMD simulations to obtain sufficient sampling of permeation events within computationally accessible timescales~\cite{liwangWaterTransportReverse2023, marioniNonequilibriumSimulationsHydraulic2026a}.
    
    Each simulation was run for $300\,\mathrm{ns}$, with transport properties evaluated over the final $150\,\mathrm{ns}$. Trajectory visualization and analysis were performed using VMD~\cite{humphreyVMDVisualMolecular1996} and MDAnalysis~\cite{gowers2016, michaud-agrawalMDAnalysisToolkitAnalysis2011}, respectively. Further details on membrane generation, the NEMD setup, and analysis protocols are provided in Supporting Information Sections~\ref{SI-sec:SI_atomistic}, \ref{SI-sec:SI_NEMD}, and \ref{SI-sec:SI_analyis}, respectively.

\section{Results}

\subsection{Comparable water fluxes but support-dependent internal transport profiles}
\label{sec:validationFlux}

    \begin{figure}[ht]
    \centering
    \includegraphics[width=\linewidth]{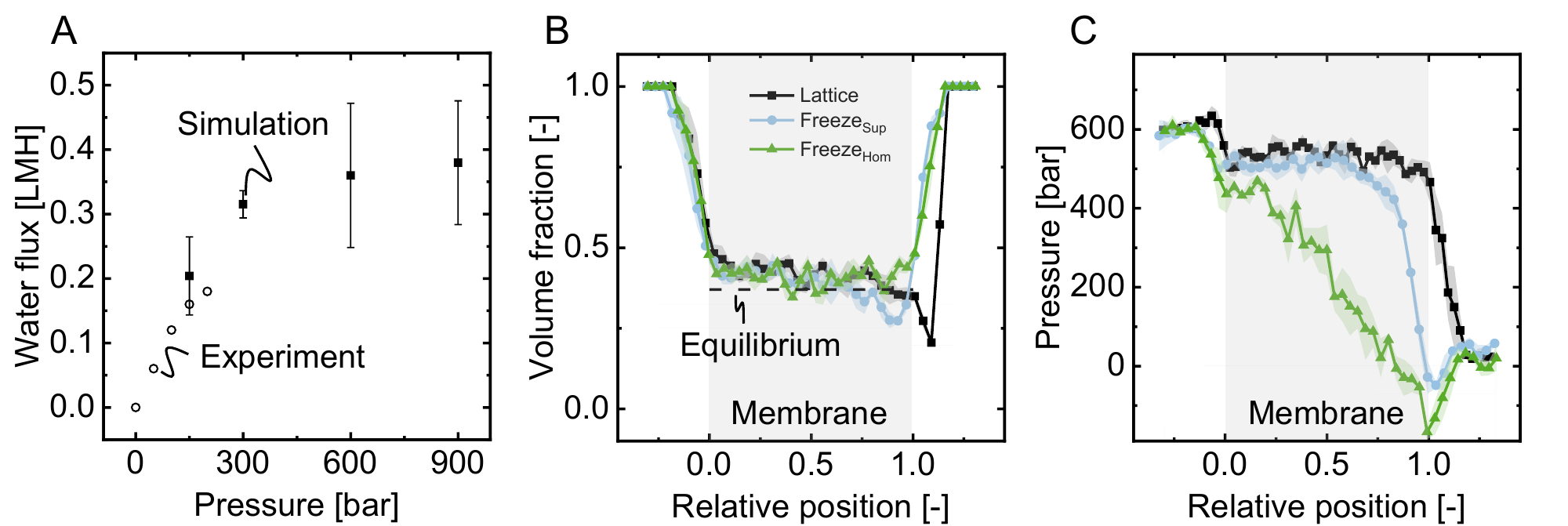}
    \caption{(\textbf{A}) Water flux as a function of the applied pressure difference for the Lattice support configuration. Simulated apparent fluxes were scaled by membrane thickness for comparison with experimental PEGDA data from Reimund et al.~\cite{reimundExperimentalObservationNonlinear2025a}. (\textbf{B}) Water volume fraction along the relative membrane position at a pressure difference of $600\,\mathrm{bar}$ for the three support strategies. Relative positions of $0$ and $1$ denote the feed- and permeate-side membrane interfaces, respectively. The dashed line marks the experimentally reported equilibrium water volume fraction~\cite{reimundExperimentalObservationNonlinear2025a}. (\textbf{C}) Pressure distribution along the relative membrane position for the same three support strategies. In all panels, averages were calculated over five independently generated membrane systems; error bars in (\textbf{A}) and shaded regions in (\textbf{B}) and (\textbf{C}) indicate the corresponding standard deviations.}
    \label{fig:fig2}
    \end{figure}
    
    The simulated system was first benchmarked against the pressure-dependent water flux reported experimentally by Reimund et al.~\cite{reimundExperimentalObservationNonlinear2025a}. Figure~\ref{fig:fig2}A shows the water flux as a function of the applied pressure difference for the Lattice configuration. To facilitate comparison with experiments, the apparent simulated water flux was scaled by membrane thickness. At $150\,\mathrm{bar}$, the resulting thickness-normalized permeability is comparable to the experimental value~\cite{reimundExperimentalObservationNonlinear2025a}, providing a useful consistency check rather than a direct quantitative validation. Such a direct comparison is limited by the large difference in membrane thickness and by the potentially greater relative contribution of interfacial resistance in the simulated system~\cite{songMolecularSimulationsWater2020}. Additional structural and transport consistency checks are provided in Supporting Information Section~\ref{SI-sec:SI_atomistic}. With increasing pressure, the flux increases sublinearly and the apparent permeability decreases, consistent with the high-pressure limitation discussed by Reimund et al.~\cite{reimundExperimentalObservationNonlinear2025a}. At $600\,\mathrm{bar}$, the water fluxes obtained with the Lattice, $\mathrm{Freeze}_{\mathrm{Sup}}$, and $\mathrm{Freeze}_{\mathrm{Hom}}$ support strategies remain similar, ranging from approximately $0.3$ to $0.4\,\mathrm{LMH}$. Thus, the three mechanical implementations generate comparable overall transport rates despite imposing different constraints on the membrane.
    
    To determine how the support strategy influences the internal membrane state during transport, Figure~\ref{fig:fig2}B compares the water volume fraction along the transport direction at a pressure difference of $600\,\mathrm{bar}$. The relative membrane coordinate ranges from $0$ at the feed-side interface to $1$ at the permeate-side interface. The membrane boundaries were defined as the positions at which the local membrane density first and last reached $90\,\%$ of the average internal membrane density. For the Lattice configuration, the water volume fraction gradually decreases from the feed to the permeate side, compatible with the concentration gradient expected under an SD description. Near the permeate-side interface, however, the water volume fraction decreases sharply, indicating local compression of the PEGDA network against the Lattice support. This deformation may be promoted by the comparatively long and flexible PEGDA chains, in contrast to the more rigid polyamide networks considered in previous simulations~\cite{marioniNonequilibriumSimulationsHydraulic2026a}. The high pressures required for NEMD sampling may further enhance this interfacial effect relative to experimental conditions, where substantially lower pressures are typically applied.
    
    For the $\mathrm{Freeze}_{\mathrm{Sup}}$ configuration, the water volume fraction also decreases moderately from the feed to the permeate side, but without the pronounced interfacial depletion observed for the Lattice configuration. By contrast, $\mathrm{Freeze}_{\mathrm{Hom}}$ exhibits an approximately uniform water volume fraction throughout the membrane. Thus, distributing the frozen atoms throughout the membrane suppresses the concentration gradient. The dashed line in Figure~\ref{fig:fig2}B indicates the experimentally reported equilibrium water volume fraction~\cite{reimundExperimentalObservationNonlinear2025a}. Its close agreement with the simulated water contents indicates that the membrane model reproduces the overall hydration state of experimental PEGDA membranes. The slight increase in water content near the feed-side interface is also consistent with previous NEMD simulations of polyamide membranes reported by Marioni et al.~\cite{marioniNonequilibriumSimulationsHydraulic2026a}. Further analysis of the internal water distribution, including void-size distributions, water profiles at additional pressure differences, and the evolution of the concentration gradient, is provided in Supporting Information Section~\ref{SI-sec:SI_WaterVolume}.
    
    Figure~\ref{fig:fig2}C compares the hydrostatic pressure distributions within the membrane for the three support strategies at $600\,\mathrm{bar}$. The pressure profiles were calculated using Gromacs-LS~\cite{vanegas_importance_2014, torres-sanchez_examining_2015, torres-sanchez_geometric_2016}. Methodological details are provided in Supporting Information Section~\ref{SI-sec:pressure}. For all three configurations, the imposed pressure difference is accurately reproduced in the bulk water reservoirs. In the Lattice configuration, the pressure remains approximately constant throughout most of the membrane and then drops sharply near the permeate-side support. This profile has an SD-like signature, with only a small hydrostatic pressure gradient through most of the membrane. The $\mathrm{Freeze}_{\mathrm{Sup}}$ configuration shows a similar overall profile, although the pressure begins to decrease near the localized frozen atoms before dropping more sharply in their immediate vicinity. In contrast, $\mathrm{Freeze}_{\mathrm{Hom}}$ exhibits an approximately linear pressure decrease across the membrane. Together with its nearly uniform water volume fraction, this profile has a distinctly more PF-like signature.
    
    The support-dependent changes in pressure and water-concentration profiles reproduce the qualitative boundary-condition sensitivity reported by Marioni et al.~\cite{marioniNonequilibriumSimulationsHydraulic2026a} for polyamide membranes. Here, the three support strategies generate clearly different SD-like and PF-like profiles in the same PEGDA membrane while maintaining comparable overall water flux. The membrane-normal stress in Supporting Information Figure~\ref{SI-fig:SI_NormalPressure} shows the same support dependence. We therefore next ask whether these macroscopic profile differences are accompanied by corresponding changes in molecular water dynamics.

\subsection{Molecular water dynamics remain similar across support strategies}
\label{sec:microscopicMobility}

    \begin{figure}[ht!]
    \centering
    \includegraphics[width=\linewidth]{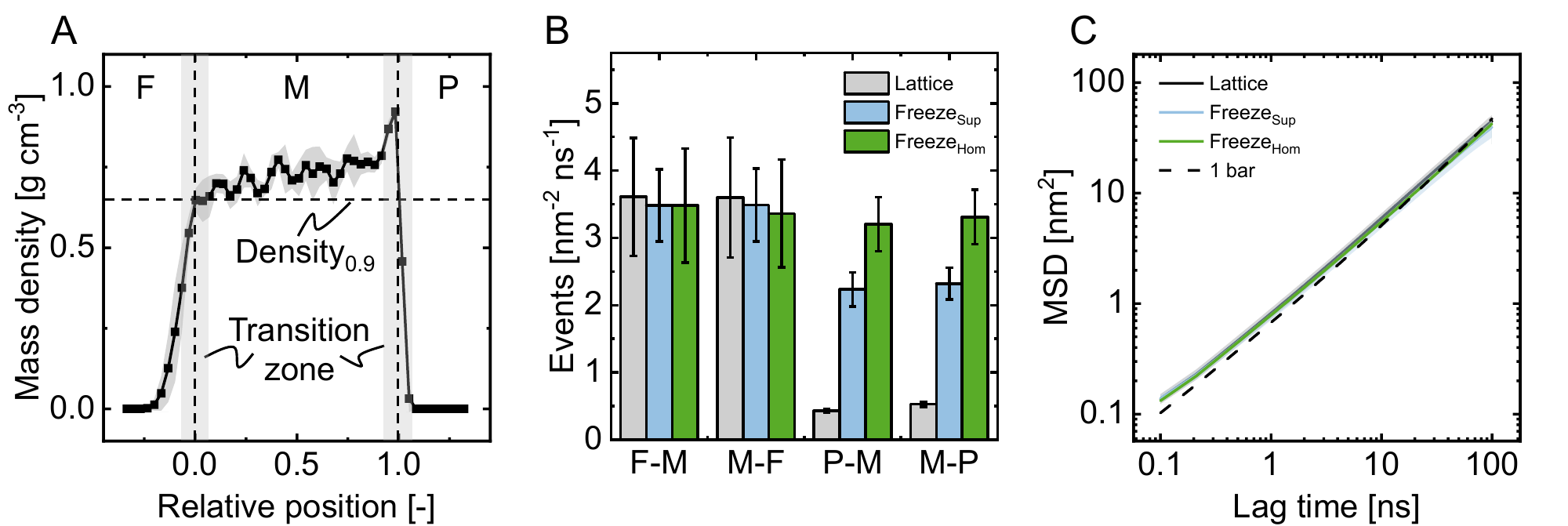}
    \caption{(\textbf{A}) Membrane mass-density profile along the relative transport coordinate for the Lattice configuration at a pressure difference of $600\,\mathrm{bar}$, illustrating the spatial regions used for the molecular-level analysis. Membrane boundaries were defined as the positions at which the local density first and last reached $90\,\%$ of the average internal membrane density. Shaded regions indicate the $1\,\mathrm{nm}$-wide interfacial transition zones. (\textbf{B}) Frequency of water partitioning events across the membrane interfaces for the three support strategies normalized by interfacial area and simulation time. Four event types were distinguished: feed to membrane (F--M), membrane to feed (M--F), permeate to membrane (P--M), and membrane to permeate (M--P). (\textbf{C}) Centered mean squared displacement (MSD) of water molecules within the membrane. The dashed line indicates the reference MSD obtained at $1\,\mathrm{bar}$ for a homogeneous membrane system.}
    \label{fig:fig3}
    \end{figure}

    We first compare interfacial water exchange and intramembrane mobility across the three support strategies. Figure~\ref{fig:fig3}A illustrates the spatial regions used for this analysis. The feed (F), membrane (M), and permeate (P) regions are shown together with the membrane mass-density profile along the relative transport coordinate. Two $1\,\mathrm{nm}$-wide interfacial transition zones were defined at the feed/membrane and membrane/permeate interfaces, centered at relative positions $0$ and $1$, respectively. For the Lattice configuration shown here, the membrane mass density increases moderately from the feed toward the permeate side, indicating progressive compaction of the polymer network along the transport direction. This trend is consistent with the corresponding decrease in water volume fraction shown in Figure~\ref{fig:fig2}B.
    
    Figure~\ref{fig:fig3}B shows the frequency of water partitioning events across the two membrane interfaces for the three support strategies at a fixed pressure difference of $600\,\mathrm{bar}$. Four event types were distinguished: feed to membrane (F--M), membrane to feed (M--F), permeate to membrane (P--M), and membrane to permeate (M--P). A partitioning event was counted when a water molecule crossed the corresponding transition zone and was subsequently detected in the adjacent region. The resulting frequencies were normalized by interfacial area and simulation time. At the feed/membrane interface, partitioning frequencies are similar for the Lattice, $\mathrm{Freeze}_{\mathrm{Sup}}$, and $\mathrm{Freeze}_{\mathrm{Hom}}$ configurations, with forward and backward transitions occurring at comparable rates.
    
    To place these interfacial fluctuations in relation to the net transport rate, the mean feed-side exchange frequency, defined as $(J_{\mathrm{F-M}}+J_{\mathrm{M-F}})/2$, was compared with the independently determined net water flux shown in Figure~\ref{fig:fig2}A. The resulting order-of-magnitude estimate is approximately six orders of magnitude larger than the net permeation flux. Thus, the net flux is a small directional imbalance relative to much larger molecular exchange in both directions. This comparison does not, by itself, distinguish SD from PF transport and should not be interpreted as a direct flux balance because the absolute partitioning frequency depends on the definition and width of the interfacial transition zone. It nevertheless shows how small the net directional transport is compared with the reversible exchange at the interface.
    
    In contrast to the similar feed-side exchange dynamics, pronounced support-dependent differences emerge at the membrane/permeate interface. The Lattice configuration exhibits the lowest partitioning frequency, followed by $\mathrm{Freeze}_{\mathrm{Sup}}$, whereas $\mathrm{Freeze}_{\mathrm{Hom}}$ shows the highest frequency, approaching the magnitude observed at the feed-side interface. This ordering mirrors the corresponding water volume-fraction profiles in Figure~\ref{fig:fig2}B. The pronounced water depletion near the permeate-side interface in the Lattice configuration reduces the local population of water molecules available for interfacial exchange, whereas this depletion is weaker for $\mathrm{Freeze}_{\mathrm{Sup}}$ and absent for $\mathrm{Freeze}_{\mathrm{Hom}}$. For the Lattice configuration, the pressure dependence of the partitioning frequencies is shown in Supporting Information Figure~\ref{SI-fig:SI_WaterPartititioning}. The permeate-side partitioning frequency decreases with increasing pressure difference, consistent with stronger local membrane compaction and the associated reduction in water content near the permeate interface. Overall, water exchange remains strongly bidirectional for all three support strategies, while the magnitude of permeate-side exchange is governed primarily by the local, support-dependent membrane structure and hydration state.
    
    Intramembrane water mobility was further quantified using the centered mean squared displacement (MSD), shown in Figure~\ref{fig:fig3}C. Centering removes the net drift along the membrane-normal $z$ direction and thereby isolates the diffusive contribution to molecular motion; the corresponding drift is analyzed separately in Supporting Information Section~\ref{SI-sec:SI_WaterDiffusivity}. In the pressure-driven simulations, the MSD curves deviate from linear behavior at longer lag times because water molecules that leave the membrane are excluded from subsequent sampling. This progressively biases the remaining population toward more slowly diffusing molecules. Diffusion coefficients were therefore extracted only from the initial diffusive regime, defined by a local slope greater than $0.9$ in the log--log representation of the MSD and by retention of at least $90\,\%$ of the initially sampled water molecules within the membrane.
    
    The resulting diffusion coefficients are similar across all support strategies: $0.88\times10^{-10}\,\mathrm{m^2\,s^{-1}}$ for Lattice, $0.81\times10^{-10}\,\mathrm{m^2\,s^{-1}}$ for $\mathrm{Freeze}_{\mathrm{Sup}}$, and $0.79\times10^{-10}\,\mathrm{m^2\,s^{-1}}$ for $\mathrm{Freeze}_{\mathrm{Hom}}$. The equilibrium reference simulation at $1\,\mathrm{bar}$ yields $0.79\times10^{-10}\,\mathrm{m^2\,s^{-1}}$, in close agreement with the experimental value reported by Reimund et al.~\cite{reimundExperimentalObservationNonlinear2025a} for the same PEGDA membrane system. Although the support strategy changes the pressure and water-content profiles substantially and modifies the local interfacial exchange near the permeate side, the diffusion coefficient within the membrane changes only slightly. Full details of the MSD analysis, including results at additional pressure differences, are provided in Supporting Information Section~\ref{SI-sec:SI_WaterDiffusivity}.

    \begin{figure}[ht]
    \centering
    \includegraphics[width=\linewidth]{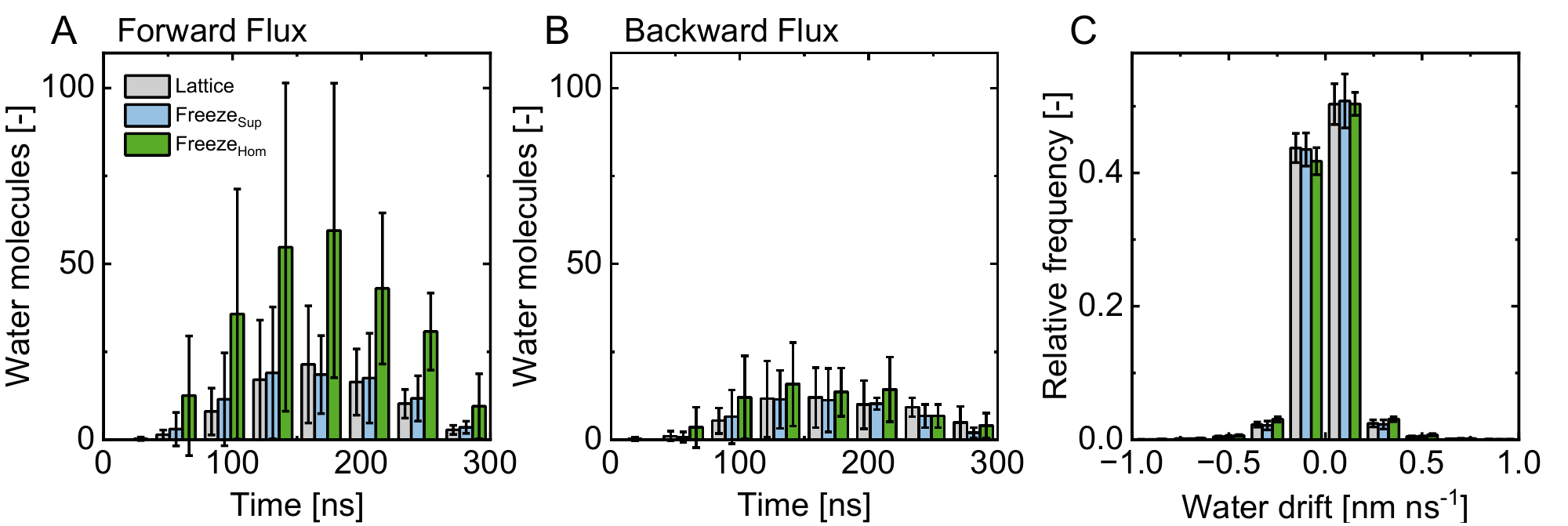}
    \caption{(\textbf{A,B}) Residence times of complete water translocations across the membrane in the forward (feed-to-permeate, A) and backward (permeate-to-feed, B) directions for the investigated support strategies. (\textbf{C}) Water drift along the membrane-normal $z$ direction, calculated as the net $z$ displacement normalized by the residence time within the membrane. Positive and negative values indicate net motion toward the permeate and feed sides, respectively.}
    \label{fig:fig4}
    \end{figure}
    
    The diffusivity analysis quantifies intramembrane mobility but does not resolve the directional bias of individual molecular trajectories. We therefore tracked complete translocation events and the net drift of individual water molecules along the transport direction. Figures~\ref{fig:fig4}A and B show the residence times of molecules that completely traversed the membrane. Full translocations occur both from feed to permeate (Figure~\ref{fig:fig4}A) and from permeate to feed (Figure~\ref{fig:fig4}B). These reverse crossings do not by themselves exclude a pressure-driven contribution to transport. They show, however, that water continues to move in both directions even when the net flux is toward the permeate side.

    Among the three support strategies, $\mathrm{Freeze}_{\mathrm{Hom}}$ exhibits the largest number of complete translocation events. This may reflect the lower degree of membrane compaction associated with this restraint scheme. The residence-time analysis is, however, subject to substantial statistical uncertainty because only a small fraction of water molecules fully traverses the membrane within the accessible simulation time. Moreover, molecules with longer residence times are less likely to complete a full crossing during the observation window and are therefore underrepresented. Consequently, the translocation events shown in Figures~\ref{fig:fig4}A and B predominantly characterize the more mobile fraction of the membrane water population. Results for additional pressure differences are provided in Supporting Information Figure~\ref{fig:SI_Residence}.
    
    To obtain a more statistically representative measure of directional transport, Figure~\ref{fig:fig4}C shows the water drift along the membrane-normal $z$ direction. Unlike the complete-translocation analysis, this metric includes all water molecules residing within the membrane, regardless of whether they traverse the full membrane thickness. The drift was calculated from the net displacement of each water molecule along $z$, normalized by its residence time within the membrane region. Most water molecules exhibit only small net drift values, consistent with local motion or temporary confinement within the polymer network, for example in poorly connected water-rich regions. A smaller fraction displays larger positive or negative drift values, producing a broad distribution around zero. All support strategies show similar broad distributions with a modest shift toward the permeate direction. The pressure difference therefore biases the molecular motion rather than making it uniformly directed.
    
    The complete-translocation and drift analyses probe different parts of the same molecular process. Complete crossings select the more mobile molecules that traverse the membrane within the simulation time, whereas the drift distribution includes the broader membrane-water population and is dominated by smaller local displacements. Neither measure changes qualitatively with the support strategy, although the pressure and concentration profiles do. Together with the similar diffusivities and the bidirectional interfacial exchange, this shows that different macroscopic profiles can occur with similar molecular water dynamics.

    \subsection{Local water correlations are short-lived and similar across supports}

    \begin{figure}[ht]
    \centering
    \includegraphics[width=\linewidth]{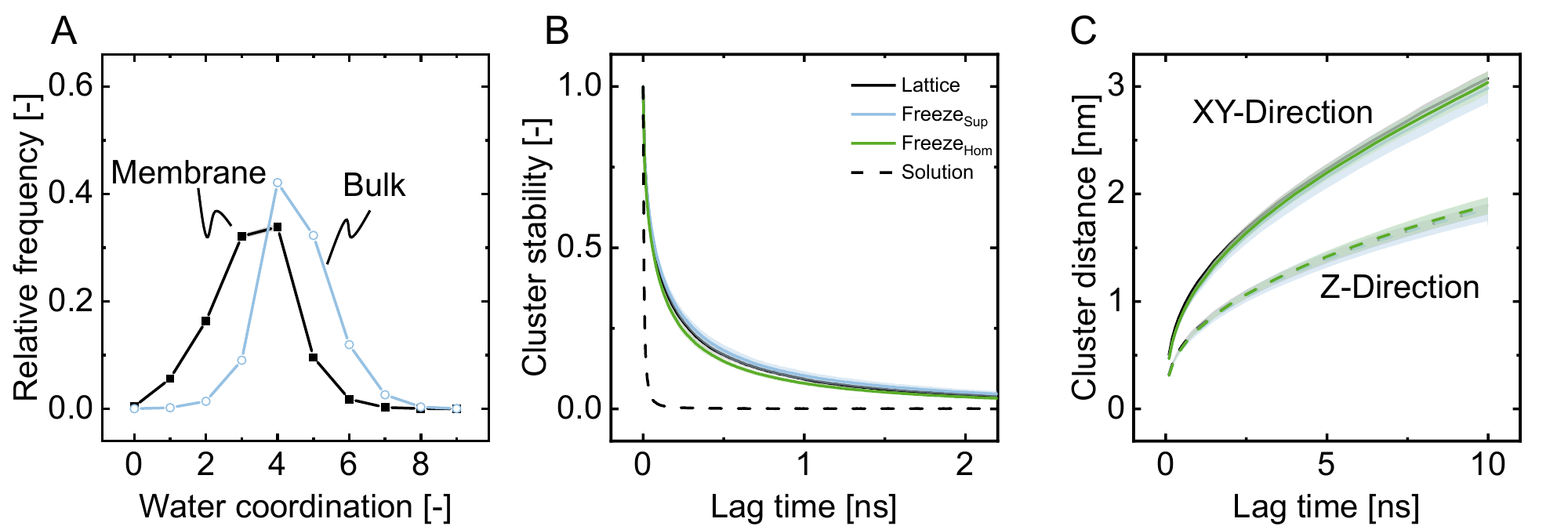}
    \caption{(\textbf{A}) Average number of nearest water neighbors in the bulk solution and within the membrane. (\textbf{B}) Survival probability of water clusters in the bulk solution and within the membrane. The survival probability quantifies the fraction of water molecules belonging to the initial neighbor shell that remain associated with the reference molecule after a given lag time, allowing for intermittent dissociation and reassociation. A value of $1$ indicates complete retention of the initial neighbor shell, whereas a value of $0$ indicates complete loss of the initially associated water molecules. (\textbf{C}) Pairwise separation between a central water molecule and the surrounding water molecules of its initial cluster as a function of lag time, resolved into the membrane-parallel ($xy$) and membrane-normal ($z$) directions.}
    \label{fig:fig5}
    \end{figure}
    
    We next examined whether neighboring water molecules move together over appreciable distances and times. Persistent collective motion would require these molecules to remain correlated, whereas rapid decorrelation would be more consistent with local diffusive motion~\cite{liwangWaterTransportReverse2023}. This analysis tests one specific picture of collective transport and does not by itself distinguish every formulation of PF from SD transport. Water clusters were defined using the first minimum of the oxygen--oxygen radial distribution function of water, as shown in Figure~\ref{fig:SI_RDF}. Figure~\ref{fig:fig5}A shows that the average number of nearest water neighbors decreases from approximately $4.3$ in bulk water to approximately $3.3$ within the membrane, consistent with confinement by the polymer network. However, the number of neighbors alone does not distinguish transport mechanisms; their lifetime is more informative.
    
    Figure~\ref{fig:fig5}B shows the survival probability of the initially defined water clusters. In bulk water, the survival probability decays rapidly, with most initial neighbor correlations lost within $100\,\mathrm{ps}$. Within the membrane, the decay is slower because molecular rearrangement is hindered by the polymer network. Nevertheless, the survival probability decreases to approximately $0.1$ within the first $1\,\mathrm{ns}$ and subsequently approaches zero, indicating that local water associations are short-lived. The Lattice, $\mathrm{Freeze}_{\mathrm{Sup}}$, and $\mathrm{Freeze}_{\mathrm{Hom}}$ configurations exhibit nearly identical decay behavior, suggesting that the support strategy has little influence on the lifetime of these associations. The corresponding pressure dependence for the Lattice configuration is shown in Supporting Information Figure~\ref{SI-fig:SI_ClusterSurvivial}A.
    
    Because cluster survival is defined using a distance cutoff, loss of a neighbor does not necessarily imply substantial molecular separation. Figure~\ref{fig:fig5}C therefore shows the pairwise separation between each reference water molecule and its initially associated neighbors, resolved into the membrane-parallel ($xy$) and membrane-normal ($z$) directions. The separation increases continuously with lag time in both directions, showing that initial neighbors progressively diffuse apart rather than merely fluctuating around the cluster cutoff. This multidirectional separation is inconsistent with long-lived water clusters moving coherently along persistent, directionally defined pathways.

    \begin{figure}[ht]
    \centering
    \includegraphics[width=\linewidth]{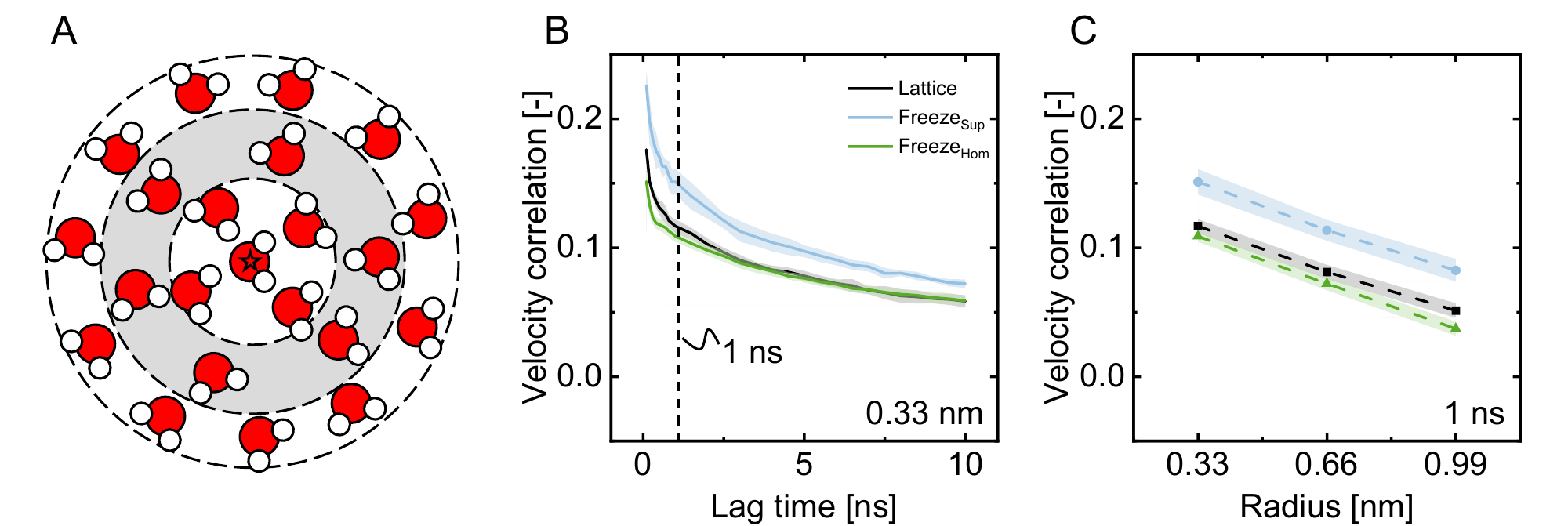}
    \caption{(\textbf{A}) Schematic definition of the coordination shells around a central water molecule, indicated by a star. Shells were defined exclusively, such that each neighboring water molecule was assigned to only one shell. In the illustrated example, the first coordination shell contains four water molecules, whereas the second contains six water molecules within the surrounding grey region. (\textbf{B}) Directional velocity correlation between a central water molecule and water molecules initially located within its first coordination shell, defined by a radius of $0.33\,\mathrm{nm}$ corresponding to the first minimum of the oxygen--oxygen radial distribution function. (\textbf{C}) Directional velocity correlation at a fixed lag time of $1\,\mathrm{ns}$ for water molecules initially located within non-overlapping coordination shells spanning $0$--$0.33$, $0.33$--$0.66$, and $0.66$--$0.99\,\mathrm{nm}$ from the central water molecule.}
    \label{fig:fig6}
    \end{figure}
    
    Directional velocity correlations were next used to determine whether initially neighboring water molecules nevertheless retain similar directions of motion after they separate. Figure~\ref{fig:fig6}A illustrates the three non-overlapping coordination shells defined around a central water molecule. The first shell extends to $0.33\,\mathrm{nm}$, corresponding to the first minimum of the oxygen--oxygen radial distribution function (Supporting Information Figure~\ref{fig:SI_RDF}), while the second and third shells span $0.33$--$0.66$ and $0.66$--$0.99\,\mathrm{nm}$, respectively. Details of the velocity-correlation analysis are provided in Supporting Information Section~\ref{SI-sec:velocity_correlation}.
    
    Water molecules were assigned to a coordination shell at the reference time $t=0$ and retained as part of that initial neighbor set throughout the analysis. Unlike the cluster-survival metric, this analysis does not require neighboring molecules to remain spatially associated but instead tests whether initially nearby molecules retain similar directions of motion. A correlation value of $1$ corresponds to perfectly aligned motion, $-1$ to oppositely directed motion, and values near $0$ to weak or negligible directional correlation.
    
    Figure~\ref{fig:fig6}B shows the directional velocity correlation between a central water molecule and molecules initially located within its first coordination shell ($r<0.33\,\mathrm{nm}$). Neighboring molecules exhibit positive correlation at short lag times, reflecting transient local coupling, but the correlation decays rapidly and approaches approximately $0.05$ at longer times. The weak residual correlation may reflect the common directional bias imposed by the chemical potential gradient across the membrane. The three support strategies show the same qualitative decay.
    
    Figure~\ref{fig:fig6}C compares the directional velocity correlation at a fixed lag time of $1\,\mathrm{ns}$ across the three coordination shells. The correlation decreases with increasing distance from the central molecule, indicating that directional coupling is strongest among immediate neighbors and weakens over larger distances.
    
    Together, Figures~\ref{fig:fig5} and~\ref{fig:fig6} show that correlations between neighboring water molecules in PEGDA are limited in both time and space. Initial neighbors exchange and diffuse apart on nanosecond timescales, while their directional velocity correlations decay rapidly and weaken with separation. The simulations therefore provide no evidence for long-lived water clusters undergoing coherent transport over extended distances. Instead, the observed behavior is consistent with predominantly local diffusive motion, and this conclusion is unchanged across the three support strategies despite their different pressure and concentration profiles.

\subsection{Implications for the solution--diffusion/pore-flow description}

    In the context of the long-standing SD--PF debate, the present results support a more nuanced interpretation of these descriptions. Early frictional and irreversible-thermodynamic treatments already allowed diffusive and hydraulic contributions to coexist in swollen polymer membranes, with their relative importance depending on hydration, frictional interactions, and membrane state~\cite{mikuleckyRelativeContributionViscous1967, yasudaDiffusiveHydraulicPermeabilities1971, daneshpajoohEquationsMembraneTransport1975}. More recent two-phase formulations have shown that SD and PF descriptions can represent complementary limits of coupled solvent–polymer transport~\cite{hegdeTwophaseModelThat2022}. The present simulations add a molecular perspective to this picture. Mechanical boundary conditions shift the calculated pressure and concentration profiles between signatures conventionally associated with SD- and PF-like transport, while molecular mobility, interfacial exchange, trajectory statistics, and short-range correlations change much less. This suggests that a change in the apparent macroscopic transport signature does not necessarily correspond to a change in the molecular mode of water transport.

\section{Conclusion}

    Mechanical support strongly influences the pressure and water-concentration profiles obtained in NEMD simulations of PEGDA membranes. The Lattice and $\mathrm{Freeze}_{\mathrm{Sup}}$ configurations produced more SD-like profiles, whereas $\mathrm{Freeze}_{\mathrm{Hom}}$ yielded a nearly uniform water concentration and an approximately linear pressure decrease. Despite these differences, the overall water flux and the molecular water dynamics remained comparatively similar across the three support strategies. Water exchange at the interfaces was strongly bidirectional, intramembrane diffusivities changed little, individual trajectories exhibited broad drift distributions with crossings in both directions, and local water correlations decayed rapidly in time and space.

    These results show that substantially different pressure and concentration profiles can arise without corresponding changes in molecular water dynamics. Such profiles alone therefore do not uniquely establish the molecular transport mechanism in mechanically restrained NEMD simulations. Mechanistic interpretation should combine spatial profiles with direct measures of molecular mobility, exchange, trajectory directionality, and intermolecular correlations.


\putbib[sample]

\end{bibunit}

\section*{CRediT authorship contribution statement}

\textbf{Jannik Mehlis:} Writing – review \& editing, Writing – original draft, Visualization, Validation, Project administration, Methodology, Investigation, Formal analysis, Data curation, Conceptualization. \textbf{Matthias Wessling:} Writing – review \& editing, Writing – original draft, Supervision, Resources, Project administration, Funding acquisition, Conceptualization.

\section*{Declaration of generative AI and AI-assisted technologies in the manuscript preparation process}

During the preparation of this work the author(s) used OpenAIs ChatGPT 5.5 to improve readability. After using this tool/service, the author(s) reviewed and edited the content as needed and take(s) full responsibility for the content of the published article.

\section*{Declaration of competing interest}

The authors declare that they have no known competing financial interests or personal relationships that could have appeared to influence the work reported in this paper.

\section*{Acknowledgements}

M.W. acknowledges DFG funding through the Gottfried Wilhelm Leibniz Award, Germany 2019 (WE 4678/12-1). Computations were performed with computing resources granted by RWTH Aachen University under project rwth2098.

\section*{Appendix A. Supplementary data}

The supplementary information includes a detailed description of the atomistic simulation, analysis protocols, and additional results not fully discussed in the main text. Code used to generate the membrane system investigated in this study is publicly available at \href{http://permalink.avt.rwth-aachen.de/?id=472286}{GitLab}.

\section*{Data availability}

Data will be made available on request.


\clearpage

\setcounter{section}{0}
\setcounter{figure}{0}
\setcounter{table}{0}
\setcounter{equation}{0}

\renewcommand{\thesection}{S\arabic{section}}
\renewcommand{\thefigure}{S\arabic{figure}}
\renewcommand{\thetable}{S\arabic{table}}
\renewcommand{\theequation}{S\arabic{equation}}

\begin{bibunit}[elsarticle-num]

\begin{center}
{\Large\bfseries Supporting Information}

\end{center}

\section{Atomistic membrane model}
\label{SI-sec:SI_atomistic}

\begin{figure}[ht]
\centering
\includegraphics[width=\linewidth]{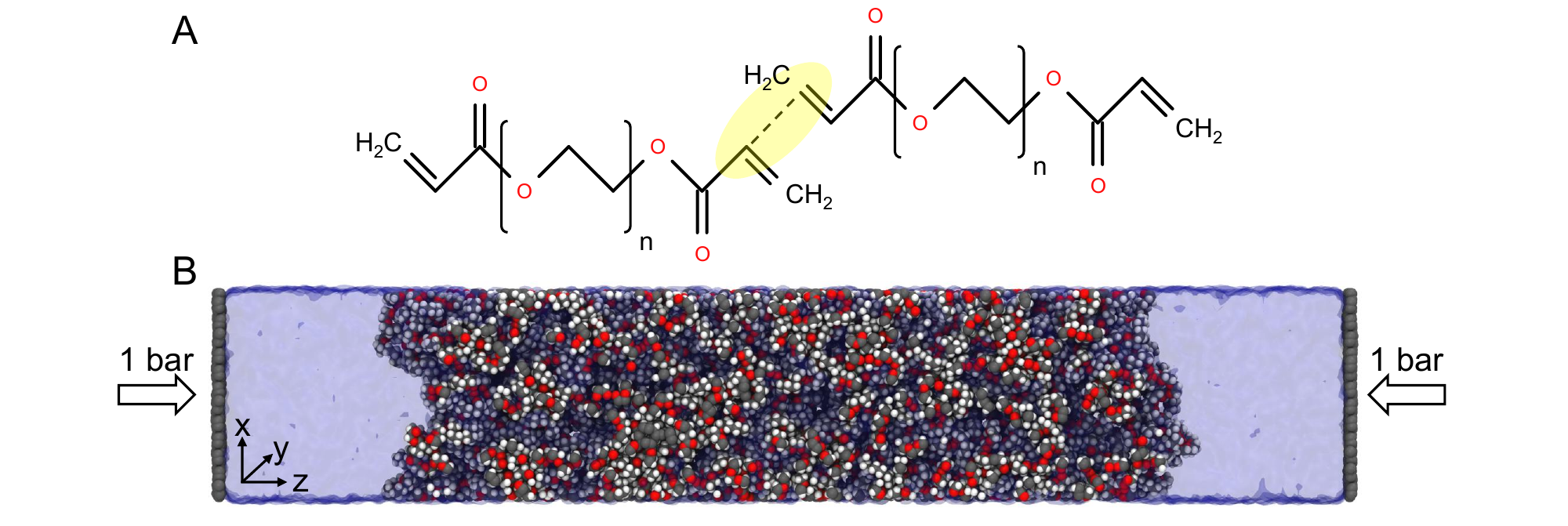}
\caption{(\textbf{A}) Schematic representation of PEGDA chain crosslinking. The crosslinking site is highlighted in yellow. The degree of polymerization is $n = 13$ in the present simulations. (\textbf{B}) Equilibration setup of the membrane slab. The membrane is positioned centrally between two graphene slabs and solvated with water on both sides. The system is then equilibrated at $1\,\mathrm{bar}$.}
\label{fig:SI_fig1}
\end{figure}

The membrane model was constructed from $300$ atomistic poly(ethylene glycol)-diacrylate (PEGDA) chains using an OPLS-AA-based~\cite{jorgensenDevelopmentTestingOPLS1996b} parameterization, following the approach of Zofchak et al.~\cite{zofchakCationPolymerInteractions2023}. The chemical structure of PEGDA and the formation of crosslinks between acrylate end groups are shown schematically in Fig.~\ref{fig:SI_fig1}A. Water was described using the TIP4P/2005 model~\cite{abascal2005general}. Each PEGDA chain contained $13$ ethylene glycol repeat units, corresponding to a molecular weight of approximately $742\,\mathrm{g\,mol^{-1}}$. The membrane water content was set to $0.5\,\mathrm{g_{water}\,g_{polymer}^{-1}}$, corresponding to a water volume fraction of approximately $37\,\mathrm{vol-\%}$, to match the system studied by Zofchak et al.~\cite{zofchakCationPolymerInteractions2023} and commonly investigated experimental PEGDA membranes~\cite{reimundExperimentalObservationNonlinear2025a, jangInfluenceWaterContent2019, jangInfluenceWaterContent2020}. During the initial equilibration, the system was confined by graphene-like walls in the direction normal to the membrane plane to maintain a slab-like geometry during equilibration and crosslinking. The wall interactions were described by a $12$--$6$ Lennard--Jones wall potential using graphene carbon parameters~\cite{marioniNonequilibriumSimulationsHydraulic2026a}. The fabrication protocol for the atomistic PEGDA membrane is supplied on \href{http://permalink.avt.rwth-aachen.de/?id=472286}{GitLab}.

The solvated PEGDA system was equilibrated following a previously established protocol~\cite{mehlisDehydrationLocalCoordination2026, mehlisDiffusiveWaterTransport2025, qiaoPolyStyrenesulfonatePoly2010}. First, the system was energy-minimized to remove unfavorable atomic contacts. Simulated annealing was then performed to relax the polymer conformations and improve chain packing, during which the temperature was reduced from $500$ to $300\,\mathrm{K}$ over $4\,\mathrm{ns}$. The system was subsequently compressed at $500\,\mathrm{bar}$ for $6\,\mathrm{ns}$ to promote hydration and densification of the membrane. To further equilibrate the hydrated system, four temperature cycles were applied. Each cycle followed the sequence $300$, $350$, $400$, $350$, and $300\,\mathrm{K}$ and lasted $14\,\mathrm{ns}$. Monitoring of the potential energy indicated that the system was equilibrated after approximately two cycles. After completion of all temperature cycles, an additional equilibration step was performed at $300\,\mathrm{K}$ and $1\,\mathrm{bar}$ for $100\,\mathrm{ns}$ before crosslinking.

A realistic PEGDA hydrogel network was generated using an in-house automated crosslinking algorithm, similar to the procedure described by Rukmani et al.~\cite{rukmaniMolecularModelingComplex2019}. Experimentally, PEGDA crosslinking proceeds through radical polymerization of the acrylate end groups~\cite{jangInfluenceWaterContent2019, jangInfluenceWaterContent2020}. FTIR analysis indicates that the synthesis procedure leads to essentially complete crosslinking, as evidenced by the absence of residual acrylate C=C bonds in the spectra~\cite{linEffectCrossLinkingGas2005}. Accordingly, crosslinking in the simulation protocol was continued until no further crosslinkable pairs were detected. Similar to the chain equilibration, the crosslinking was performed between two graphene walls to produce a slab membrane and prohibit crosslinking bonds across the boundary condition in the z-direction (see coordinate system in Figure~\ref{fig:SI_fig1}B). Possible crosslinks were identified based on the distance between reactive acrylate carbon atoms, which were considered crosslinkable when separated by less than $0.7\,\mathrm{nm}$. After each crosslinking iteration, new covalent bonds were introduced between reactive acrylate groups, and the local bonded topology was updated accordingly. This process is illustrated schematically in the yellow-highlighted region of Figure~\ref{fig:SI_fig1}A. The system was then energy-minimized and briefly equilibrated to relax local structural distortions introduced by the newly formed bonds. To limit such distortions and allow sufficient relaxation of the evolving network, a maximum of four new bonds was introduced per crosslinking iteration. This iterative procedure was repeated until no additional crosslinkable pairs were found. The final crosslinking degree exceeded $99\,\%$, consistent with the experimentally expected near-complete conversion of acrylate groups. The crosslinking degree was calculated as the fraction of reacted diacrylate carbons relative to the total number of initially crosslinkable diacrylate carbons. After the final crosslinking step, only a small number of partially reacted acrylate groups remained. In the experimental system, radical polymerization can terminate through recombination or oxygen inhibition~\cite{linEffectCrossLinkingGas2005}. For simplicity, unreacted radical sites were capped with hydrogen atoms after completion of the crosslinking procedure. This finalization step removes chemically unstable radical centers and yields a closed-shell atomistic topology suitable for classical molecular dynamics simulations.

\enlargethispage{\baselineskip}

The resulting atomistic membrane model was compared with several independent structural and transport properties. The membrane water volume fraction is consistent with experimentally reported values for comparable PEGDA membranes and with previous atomistic studies~\cite{zofchakCationPolymerInteractions2023, reimundExperimentalObservationNonlinear2025a, jangInfluenceWaterContent2019, jangInfluenceWaterContent2020}. The model also employs the same OPLS-AA-based PEGDA parameterization as Zofchak et al.~\cite{zofchakCationPolymerInteractions2023} and yields water self-diffusivities consistent with experimental values, as discussed in Supporting Information Section~\ref{SI-sec:SI_WaterDiffusivity}. The void-size analysis in Section~\ref{SI-sec:SI_WaterVolume} gives void-space characteristics comparable to those reported by Zofchak et al.~\cite{zofchakCationPolymerInteractions2023}. In addition, the water permeability obtained from the pressure-driven NEMD simulations is comparable to the experimental permeability reported for PEGDA membranes, as shown in Figure~2A of the main text. These comparisons provide structural and transport consistency checks for the atomistic PEGDA membrane model.

\section{Slab equilibration and NEMD simulation}
\label{SI-sec:SI_NEMD}

After crosslinking, the PEGDA membrane system was extended in the $z$-direction (see coordinate system in Figure~\ref{fig:SI_fig1}B) for both sorption and nonequilibrium molecular dynamics (NEMD) simulations. The crosslinked membrane was positioned centrally between two graphene pistons, and water compartments were added on both sides of the membrane, as illustrated in Figure~\ref{fig:SI_fig1}B. In the sorption simulations, the graphene pistons were used to equilibrate the hydrated membrane at an applied pressure of $1\,\mathrm{bar}$, following the procedure of Marioni et al.~\cite{marioniNonequilibriumSimulationsHydraulic2026a}. This equilibration was performed for $100\,\mathrm{ns}$ to allow the membrane thickness, water distribution, and polymer conformations to relax in the final slab geometry. Equilibration was considered complete once the water volume fraction in the central membrane region remained stable over time.

After equilibration, one of three mechanical support strategies was introduced for the subsequent NEMD simulations: a graphene support lattice on the permeate side (Lattice), frozen membrane atoms localized near the permeate-side interface ($\mathrm{Freeze}_{\mathrm{Sup}}$), or homogeneously distributed frozen membrane atoms throughout the membrane ($\mathrm{Freeze}_{\mathrm{Hom}}$). The support lattice interacted only with membrane atoms and had no interaction parameters with water, thereby providing mechanical support without directly obstructing water transport~\cite{marioniNonequilibriumSimulationsHydraulic2026a}. Following Marioni et al.~\cite{marioniNonequilibriumSimulationsHydraulic2026a}, the frozen-atom support schemes used a restraint density of $25$ frozen atoms per nm of membrane thickness. For $\mathrm{Freeze}_{\mathrm{Sup}}$, these atoms were selected near the permeate-side interface, whereas for $\mathrm{Freeze}_{\mathrm{Hom}}$ they were distributed homogeneously throughout the membrane.

The molecular dynamics settings were otherwise identical for the sorption and pressure-driven NEMD simulations. All molecular dynamics simulations were performed with GROMACS~2024.4~\cite{abrahamGROMACSHighPerformance2015, pallTacklingExascaleSoftware2015}. Temperature was controlled at $300\,\mathrm{K}$ using the Nosé--Hoover thermostat with a coupling time of $\tau = 1\,\mathrm{ps}$. Water and membrane atoms were coupled to separate temperature baths, while the graphene pistons were excluded from temperature coupling. Neighbor searching was performed up to $1.2\,\mathrm{nm}$ and updated every $10$ steps. Lennard--Jones interactions were truncated at the same cutoff distance, and long-range electrostatic interactions were calculated using the particle mesh Ewald method. Five independently generated membrane systems were prepared using the same protocol to assess the influence of initial chain packing and crosslinking topology on the calculated transport properties.

\section{Analyzing water transport}
\label{SI-sec:SI_analyis}

All trajectory analyses were performed on the final $150\,\mathrm{ns}$ of each $300\,\mathrm{ns}$ NEMD simulation using MDAnalysis~\cite{gowers2016, michaud-agrawalMDAnalysisToolkitAnalysis2011}. The initial $150\,\mathrm{ns}$ were excluded as an equilibration period based on the establishment of a linear water flux response, shown below in Figure~\ref{fig:SI_WaterPerm}, and fully developed water concentration profiles within the membrane region, as discussed in Section~\ref{SI-sec:SI_WaterVolume}. Unless stated otherwise, reported values represent averages over five independently generated PEGDA membrane systems, with standard deviations calculated across these independent systems.

\subsection{Water flux calculation}
\label{SI-sec:flux}

\begin{figure}[ht]
\centering
\includegraphics[width=\linewidth]{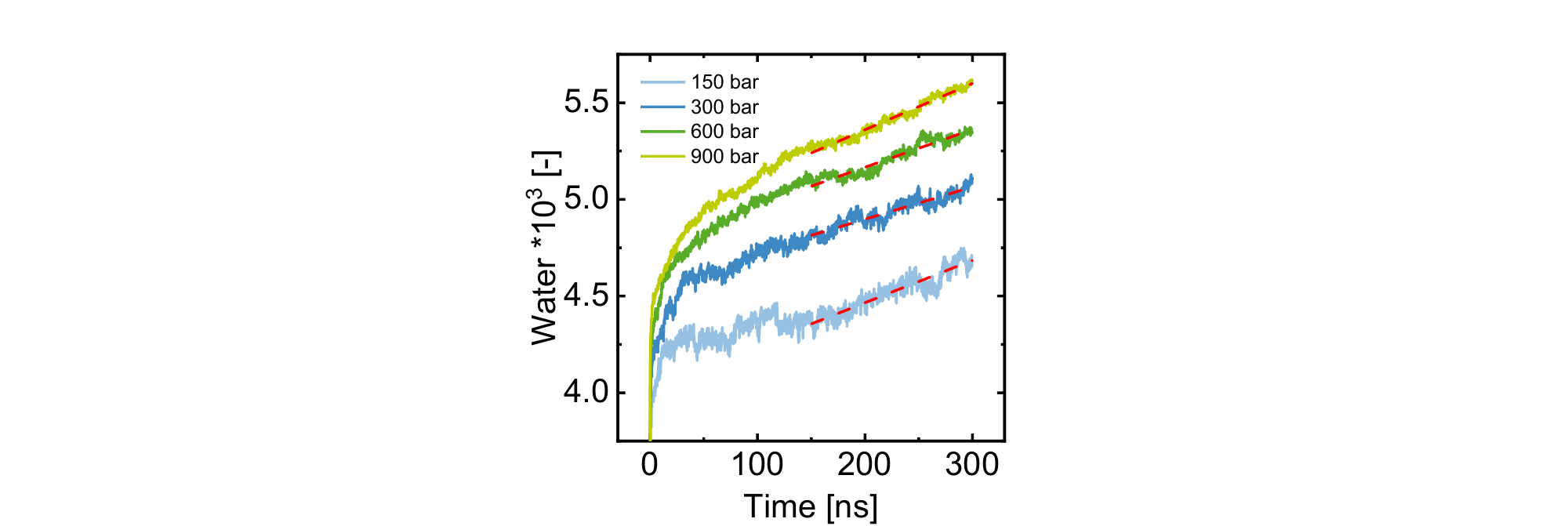}
\caption{Number of water molecules in the permeate, defined as molecules with a z-position above the support lattice structure, as a function of simulation time. The red dashed lines show linear fits to the final $150\,\mathrm{ns}$ of each simulation. The different colors indicate separate pressure differences. One representative membrane system with the Lattice support is shown.}
\label{fig:SI_WaterPerm}
\end{figure}

The water flux was calculated from the cumulative number of water molecules entering the permeate compartment during the sampling period, as shown for one membrane system in Figure~\ref{fig:SI_WaterPerm}. A linear function was fitted to the cumulative number of permeated water molecules over this interval, as indicated by the red dashed lines. The resulting slope was normalized by the membrane cross-sectional area to obtain the apparent water flux. To enable comparison with experimental PEGDA membranes, the simulated flux was further scaled by the membrane thickness. The simulated membrane was approximately $12\,\mathrm{nm}$ thick, whereas the experimental PEGDA membrane had a thickness of $440\,\mathrm{\mu m}$~\cite{reimundExperimentalObservationNonlinear2025a}. However, although thickness-based scaling is commonly used, it may overestimate the contribution of interfacial resistance in the simulated membrane~\cite{songMolecularSimulationsWater2020}.

\subsection{Local pressure profiles}
\label{SI-sec:pressure}

Local pressure profiles were calculated along the transport direction using GROMACS-LS. The spatial Irving--Kirkwood--Noll stress tensor was evaluated using the default covariant central force decomposition (cCFD) for multibody interactions~\cite{vanegas_importance_2014,torres-sanchez_examining_2015,torres-sanchez_geometric_2016}. All supported configurational contributions, including bonded, nonbonded, and constraint contributions, as well as the kinetic contribution, were included. The simulation box was divided into slabs with a thickness of approximately $0.5\,\mathrm{nm}$ normal to the membrane plane (z-direction), and the local stress tensor was averaged over the lateral x- and y-directions. The pressure profile shown in the main text corresponds to the local scalar pressure, calculated as $P=(P_{xx}+P_{yy}+P_{zz})/3$.

\subsection{Water partitioning events}
\label{SI-sec:partitioning}

Water partitioning events were identified by tracking individual water molecules across the interfacial transition zones. Four event types were distinguished: feed-to-membrane (F--M), membrane-to-feed (M--F), permeate-to-membrane (P--M), and membrane-to-permeate (M--P). An event was counted when a water molecule crossed the corresponding transition zone and was subsequently detected in the adjacent region. For example, an F--M event was assigned when a water molecule moved from the feed region through the feed/membrane transition zone and entered the membrane region. Event frequencies were normalized by the membrane cross-sectional area and the simulation time, yielding units of $\mathrm{nm^{-2}\,ns^{-1}}$.

\subsection{Water drift inside the membrane}
\label{SI-sec:drift}

The drift of individual water molecules inside the membrane was calculated from their net displacement along the transport direction during their residence time within the membrane region. For each water molecule, the position at membrane entry was subtracted from the position at membrane exit or, for molecules that did not leave the membrane during the sampling window, from the final position within the trajectory segment. The resulting displacement was divided by the corresponding residence time to obtain an average drift velocity along the transport direction. Positive drift values indicate net motion toward the permeate side, whereas negative values indicate net motion toward the feed side.

\subsection{Water--water coordination and cluster definition}
\label{SI-sec:clusters}

\begin{figure}[h]
\centering
\includegraphics[width=\linewidth]{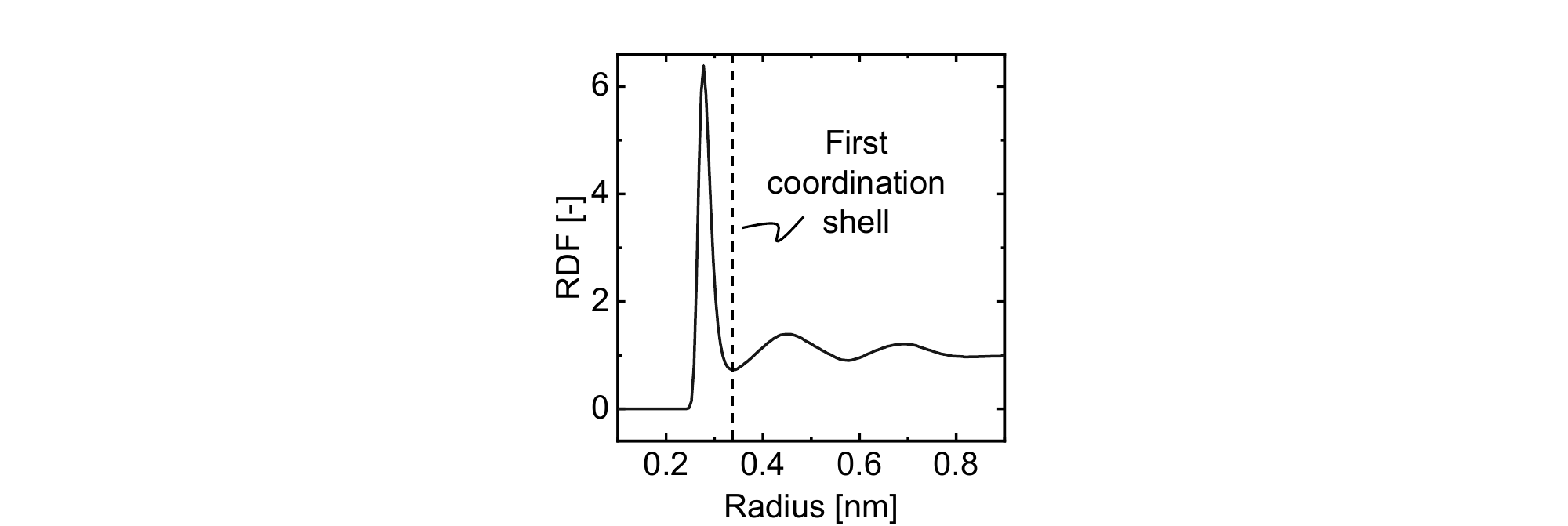}
\caption{Radial distribution function of water oxygen--oxygen distances inside the membrane. The first minimum is indicated by the dashed vertical line at $0.33\,\mathrm{nm}$.}
\label{fig:SI_RDF}
\end{figure}

Local water coordination was determined from the oxygen--oxygen radial distribution function of water (see Figure~\ref{fig:SI_RDF}). Water molecules were considered nearest neighbors if their oxygen--oxygen distance was smaller than the first minimum of the radial distribution function. This cutoff was used consistently for water molecules in the bulk solution and inside the membrane. The average number of nearest water neighbors was calculated for each water molecule and then averaged over time and over independent membrane systems.

\subsection{Water-cluster survival probability}
\label{SI-sec:cluster_survival}

The temporal stability of local water clusters was quantified using a survival probability analysis from MDAnalysis~\cite{gowers2016, michaud-agrawalMDAnalysisToolkitAnalysis2011}. For each reference water molecule, the initially neighboring water molecules were identified according to the oxygen--oxygen distance cutoff. The survival probability was then calculated as the fraction of these initially associated neighbors that remained associated after a given lag time. Intermittent dissociation and reassociation were allowed, meaning that a water molecule that temporarily left the neighbor shell but later re-entered it was still counted as correlated. A survival probability of $1$ therefore corresponds to complete retention of the initial neighbor shell, whereas a value of $0$ indicates complete loss of the initially defined local cluster.

\subsection{Directional velocity correlation of neighboring water molecules}
\label{SI-sec:velocity_correlation}

Directional velocity correlations were calculated for initially associated water molecules following an approach similar to a previous NEMD study~\cite{gochhayatWaterFlowsCollectively2026}. For each water molecule and its initial nearest neighbors, the velocity vectors were compared over increasing lag times. The correlation was calculated from the normalized scalar product of the velocity vectors, where a value of $1$ indicates perfectly aligned motion, $-1$ indicates motion in opposite directions, and $0$ indicates uncorrelated motion. The resulting correlation functions were averaged over all initially associated water pairs, over the sampling interval, and over the independent membrane systems. Unlike the water-cluster survival analysis described above, this approach does not require initially associated water molecules to remain in close proximity. The molecules may therefore diffuse apart while still exhibiting a positive directional velocity correlation.

\subsection{Void size distribution}

Void-size distributions were calculated using the open-source software PoreBlazer v4.0~\cite{sarkisovMaterialsInformaticsPoreBlazer2020}. Owing to the high computational cost, five equidistant snapshots were extracted from the final $50\,\mathrm{ns}$ of each simulation trajectory. All water molecules were removed from these snapshots to obtain simulation boxes containing only the polymer matrix and the associated void space.

\subsection{Water diffusivity}
\label{SI-sec:diffusivity_method}

Water self-diffusion coefficients were determined from mean-squared displacement (MSD) calculations using the MSD analysis module implemented in MDAnalysis~\cite{gowers2016, michaud-agrawalMDAnalysisToolkitAnalysis2011}. Following the procedure used by Zofchak et al.~\cite{zofchakCationPolymerInteractions2023}, diffusion coefficients were extracted only from time intervals exhibiting approximately Fickian behavior, defined by a logarithmic MSD slope of $0.9 < \mathrm{d}\log(\mathrm{MSD})/\mathrm{d}\log(t) < 1.1$. For the homogeneous membrane systems without a slab configuration, the MSD was calculated directly for all water molecules. For the pressure-driven slab simulations discussed in the main text, the analysis was restricted to water molecules initially located within the central $50\,\%$ of the membrane thickness, excluding the outer $25\,\%$ on either side to minimize interfacial effects. Molecules were removed from the MSD ensemble once they left the membrane region, thereby preventing their subsequent bulk-water motion from artificially increasing the calculated diffusivity. To limit the opposing bias caused by preferentially removing highly mobile molecules, MSD sampling was terminated once the surviving population decreased below $90\,\%$ of the initially selected molecules.

\section{Membrane characterization}
\label{SI-sec:SI_WaterVolume}

\begin{figure}[ht]
\centering
\includegraphics[width=\linewidth]{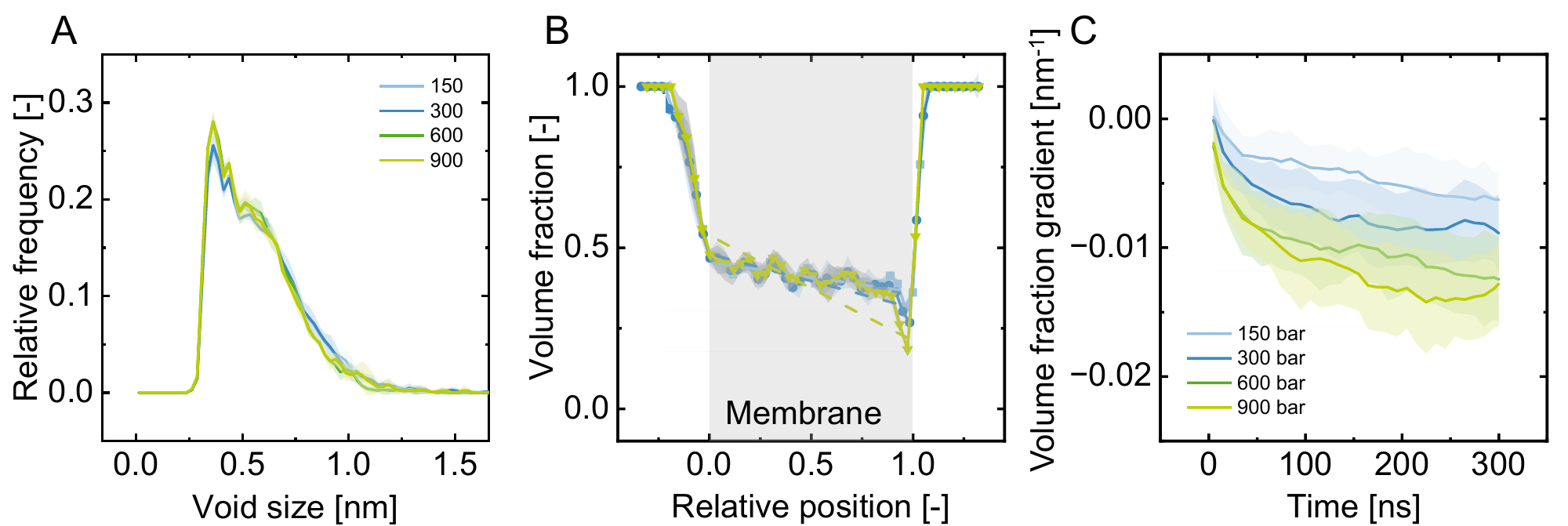}
\caption{(\textbf{A}) Void-size distributions within the membrane region at different pressure differences. (\textbf{B}) Water volume fraction as a function of the relative position across the membrane, where $0$ and $1$ denote the feed- and permeate-side interfaces, respectively. The lines correspond to pressure differences of $150$, $300$, and $900\,\mathrm{bar}$; the $600\,\mathrm{bar}$ case is shown in Figure~2B of the main text. Dashed lines indicate linear fits to the water volume fraction within the membrane region. (\textbf{C}) Time evolution of the water volume fraction gradient within the membrane, evaluated in $10\,\mathrm{ns}$ blocks.}
\label{fig:SI_WaterVolume}
\end{figure}

This section examines the void-size distribution, water volume fraction, and temporal evolution of the water concentration gradient within the membrane. All results in this section are produced with the Lattice support structure. Figure~\ref{fig:SI_WaterVolume}A shows the void-size distribution within the central membrane region at different pressure differences. The distributions are largely unaffected by pressure. Only a slight increase in the population of smaller voids centered at approximately $3.6\,\mathrm{nm}$ is observed, which may be associated with the membrane compaction near the support lattice discussed in the main text. Overall, the average void size is approximately $5.6\,\mathrm{nm}$, in good agreement with the previous simulations of Zofchak et al.~\cite{zofchakCationPolymerInteractions2023}.

Figure~\ref{fig:SI_WaterVolume}B shows the water volume fraction as a function of the relative membrane position for the pressure differences not presented in Figure~2B of the main text. For all pressure differences, the water volume fraction decreases toward the permeate side, producing a concentration gradient across the membrane. The slight membrane compaction at the position of the support lattice, corresponding to a relative position of one, is also visible and appears to become more pronounced with increasing pressure difference. The dashed lines indicate linear fits to the water volume fraction within the membrane region.

Figure~\ref{fig:SI_WaterVolume}C shows the temporal evolution of the slopes obtained from the linear fits in Figure~\ref{fig:SI_WaterVolume}B. The slopes were calculated from averages over consecutive $10\,\mathrm{ns}$ blocks. For all pressure differences, the magnitude of the slope decreases rapidly during the first $100\,\mathrm{ns}$ and subsequently begins to level off. This relaxation appears to be slowest for the $150\,\mathrm{bar}$ case. Nevertheless, the concentration gradient changes only slightly after $100\,\mathrm{ns}$, supporting the previously selected equilibration period of $150\,\mathrm{ns}$. As shown in Figure~\ref{fig:SI_WaterVolume}B, the fitted slope is influenced by membrane compaction near the support lattice. When the analysis is restricted to the central membrane region, the slopes of the water volume fraction profiles are more similar across the investigated pressure differences. This behavior is consistent with Figure~2A of the main text, which shows only a minor increase in water flux above $300\,\mathrm{bar}$.

\section{Normal pressure}

\begin{figure}[ht]
\centering
\includegraphics[width=\linewidth]{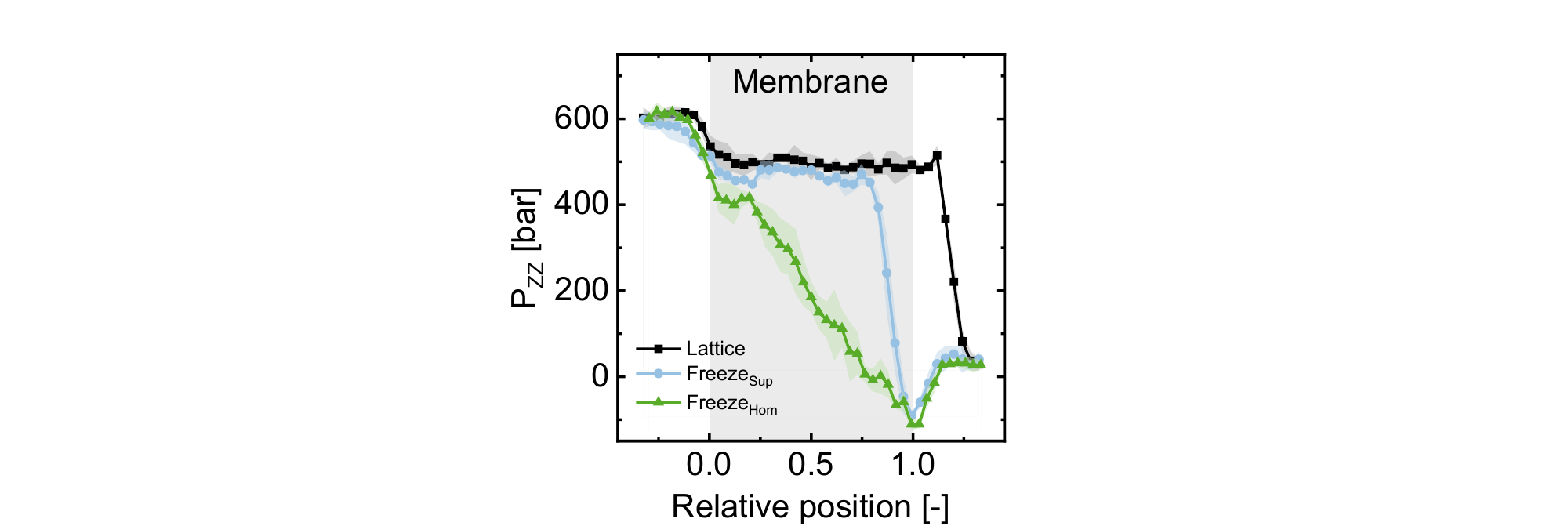}
\caption{Normal pressure distribution along the relative membrane position at a pressure difference of $600\,\mathrm{bar}$ for the three investigated membrane restraints.}
\label{SI-fig:SI_NormalPressure}
\end{figure}

\clearpage

\section{Water partition events}
\label{SI-sec:PartitionEvents}

\begin{figure}[ht]
\centering
\includegraphics[width=\linewidth]{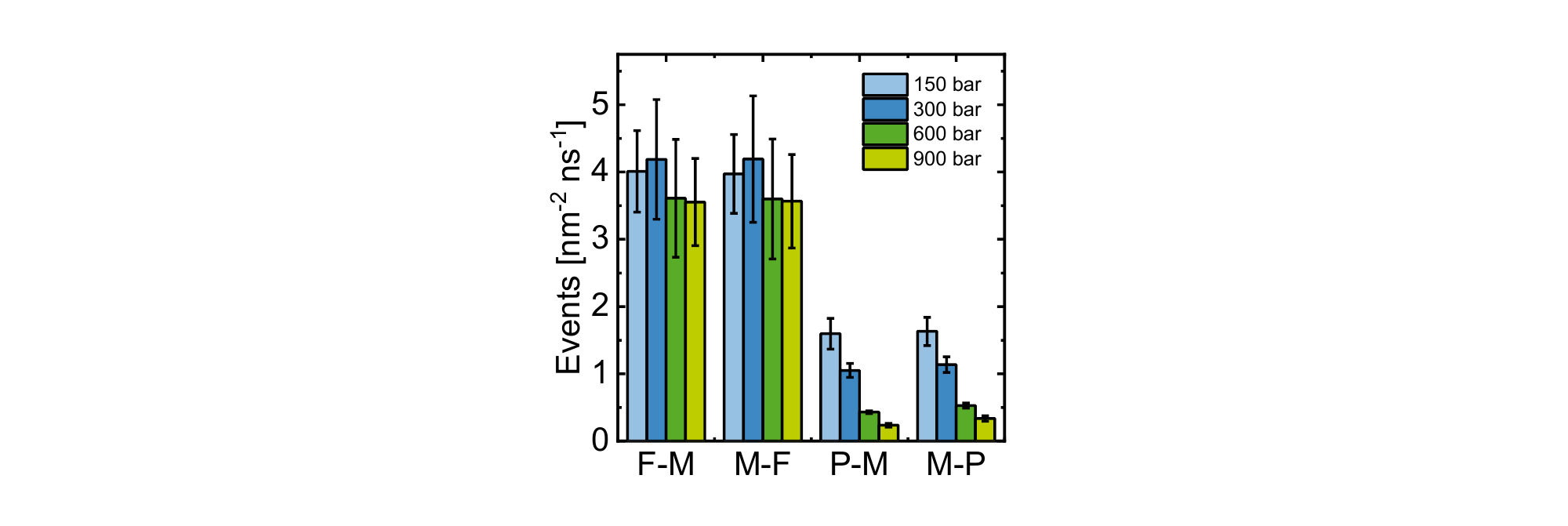}
\caption{Frequency of water partitioning events across the membrane interfaces for varying pressure gradients with the Lattice support structure, normalized by interfacial area and simulation time. Four event types were distinguished: feed to membrane (F--M), membrane to feed (M--F), permeate to membrane (P--M), and membrane to permeate (M--P).}
\label{SI-fig:SI_WaterPartititioning}
\end{figure}

\section{Water diffusivity analysis}
\label{SI-sec:SI_WaterDiffusivity}

\begin{figure}[ht]
\centering
\includegraphics[width=\linewidth]{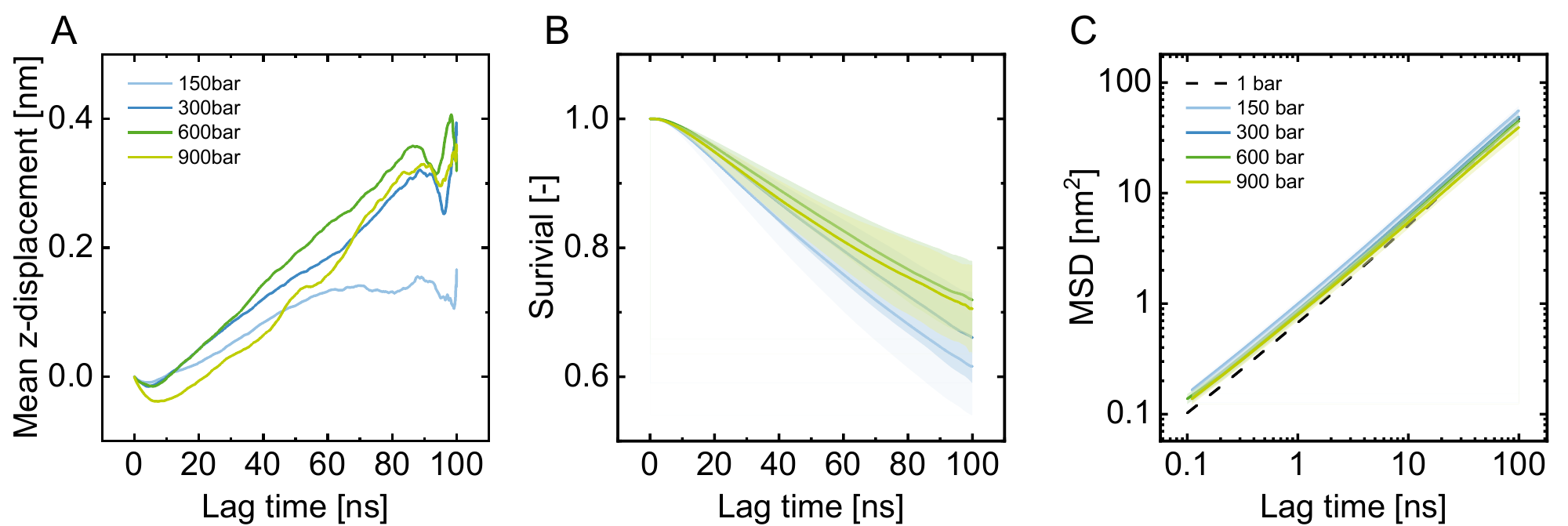}
\caption{(\textbf{A}) Mean water displacement in the $z$-direction for the different applied pressure differences. (\textbf{B}) Survival fraction of water molecules initially located within the membrane region. (\textbf{C}) Centered mean squared displacement for all investigated pressure differences. The dashed line shows the reference mean squared displacement of water in a fully crosslinked (3-D periodic) membrane at $1\,\mathrm{bar}$, obtained using a bulk membrane configuration rather than the slab geometry.}
\label{fig:SI_Diffusivity}
\end{figure}

The water self-diffusivity was calculated as described in Section~\ref{SI-sec:diffusivity_method}. Figure~\ref{fig:SI_Diffusivity}A shows the mean displacement of water molecules in the membrane-normal $z$-direction as a function of lag time for the different applied pressure differences. For clarity, the standard deviations across the independent simulations are not shown. The variation between individual runs was comparatively large, consistent with the substantial uncertainties in the directly measured water fluxes shown in Figure~2A of the main text. The mean displacement increases between the $150\,\mathrm{bar}$ case and the simulations at higher pressure differences, whereas the differences among $300$, $600$, and $900\,\mathrm{bar}$ are comparatively small. This behavior is consistent with the progressively weaker increase in water flux at high pressure differences observed in Figure~2A of the main text. At long lag times, particularly beyond approximately $50\,\mathrm{ns}$, the mean displacement also deviates from an approximately linear increase. This behavior primarily reflects the finite spatial extent of the membrane and the resulting limitations of displacement analyses in the slab geometry.

As described in Section~\ref{SI-sec:diffusivity_method}, water molecules located within the central membrane region during the final $100\,\mathrm{ns}$ of each simulation were selected for the analysis. Conventional MSD calculations are typically performed in fully periodic, spatially homogeneous systems, in which molecules remain within an equivalent environment even over long displacements. In the present slab geometry, however, water molecules eventually leave the membrane and enter the adjacent bulk-water compartments, where their dynamics differ substantially from those within the polymer network. Including such molecules after they have left the membrane would therefore artificially increase the apparent membrane diffusivity. Molecules were consequently excluded from the MSD ensemble once they exited the membrane region. The resulting survival fraction of the initially selected membrane water molecules is shown in Figure~\ref{fig:SI_Diffusivity}B. For all applied pressure differences, the survival fraction decreases approximately linearly with lag time, with the most rapid decrease observed for the $150\,\mathrm{bar}$ case. As detailed in Section~\ref{SI-sec:diffusivity_method}, the MSD analysis was restricted to lag times for which at least $90\,\%$ of the initially selected molecules remained within the membrane. At lower survival fractions, the preferential loss of highly mobile molecules increasingly biases the remaining ensemble toward slower-moving water, leading to an underestimation of the diffusivity and a deviation of the MSD slope from linearity.

Figure~\ref{fig:SI_Diffusivity}C shows the centered MSD in the membrane-normal direction. Centering removes the contribution of the net pressure-driven displacement, thereby separating fluctuations around the mean motion from the systematic drift. The molecular displacement in the membrane-normal direction can be decomposed into a stochastic and a directional contribution according to

\begin{equation}
\underbrace{\left\langle \Delta z^2(t) \right\rangle}_{\text{raw MSD}} = 
\underbrace{
\left\langle
\left[
\Delta z(t)-\left\langle\Delta z(t)\right\rangle
\right]^2
\right\rangle
}_{\text{centered MSD}}
+ 
\underbrace{
\left\langle\Delta z(t)\right\rangle^2
}_{\text{drift contribution}} .
\label{eq:msd_decomposition}
\end{equation}

Here, the first term represents the centered MSD and therefore the stochastic fluctuations around the mean molecular motion, whereas the second term originates from the systematic displacement in the direction of permeation. The mean displacement, $\langle\Delta z(t)\rangle$, is shown in Figure~\ref{fig:SI_Diffusivity}A, while the corresponding centered MSD is shown in Figure~\ref{fig:SI_Diffusivity}C. This decomposition allows the pressure-driven drift to be separated from the random molecular motion used to determine the water self-diffusivity. For reference, the MSD obtained from a homogeneous, fully periodic PEGDA membrane equilibrated at $1\,\mathrm{bar}$ is included as a dashed line in Figure~\ref{fig:SI_Diffusivity}C. In the diffusive regime, the MSD follows the Einstein relation

\begin{equation}
\mathrm{MSD}(t)=2dDt,
\label{eq:einstein_diffusion}
\end{equation}

where $d$ is the dimensionality of the displacement considered and $D$ is the corresponding self-diffusion coefficient.

The homogeneous $1\,\mathrm{bar}$ reference system exhibits an approximately linear MSD in the log--log representation, consistent with normal diffusive behavior. The resulting water self-diffusion coefficient of approximately $0.79\times10^{-10}\,\mathrm{m^2\,s^{-1}}$ agrees well with previously reported experimental water diffusivities in PEGDA hydrogels by Reimund et al.~\cite{reimundExperimentalObservationNonlinear2025a}, providing an experimental consistency check for the simulated membrane. For the pressure-driven simulations, the self-diffusion coefficients were $1.09\times10^{-10}$, $9.67\times10^{-11}$, $8.81\times10^{-11}$, and $7.81\times10^{-11}\,\mathrm{m^2\,s^{-1}}$ at pressure differences of $150$, $300$, $600$, and $900\,\mathrm{bar}$, respectively. The moderate decrease in $D$ with increasing pressure difference may reflect slight compaction of the polymer network at elevated pressures, which reduces the local free volume available for water motion. With increasing lag time, the MSD curves of the pressure-driven systems progressively deviate from their initial diffusive behavior. This deviation coincides with the decreasing survival fraction shown in Figure~\ref{fig:SI_Diffusivity}B and further illustrates the limitations of extracting long-time diffusivities in the finite slab geometry. In particular, preferential removal of molecules that traverse the membrane most rapidly leaves an increasingly slow subpopulation in the MSD ensemble. The diffusion coefficients reported here were therefore determined only over lag times satisfying both the diffusive-regime and survival-fraction criteria described above.

\section{Individual water pathways}
\label{SI-sec:SI_residenceTime}

\begin{figure}[ht]
\centering
\includegraphics[width=\linewidth]{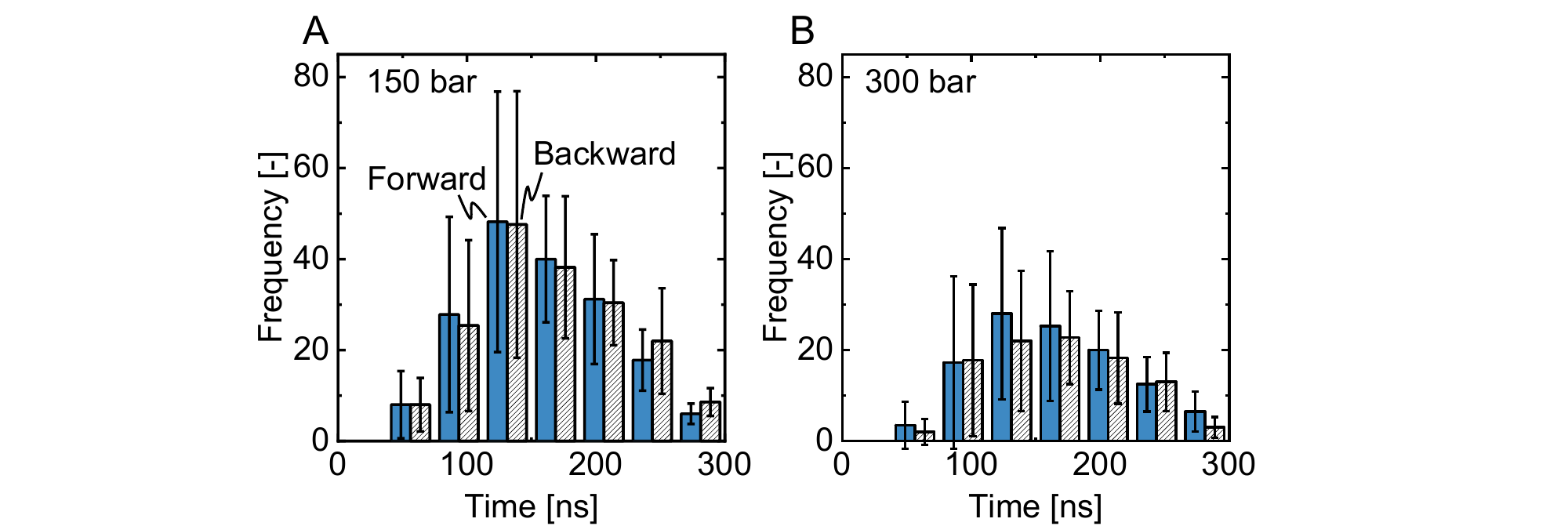}
\includegraphics[width=\linewidth]{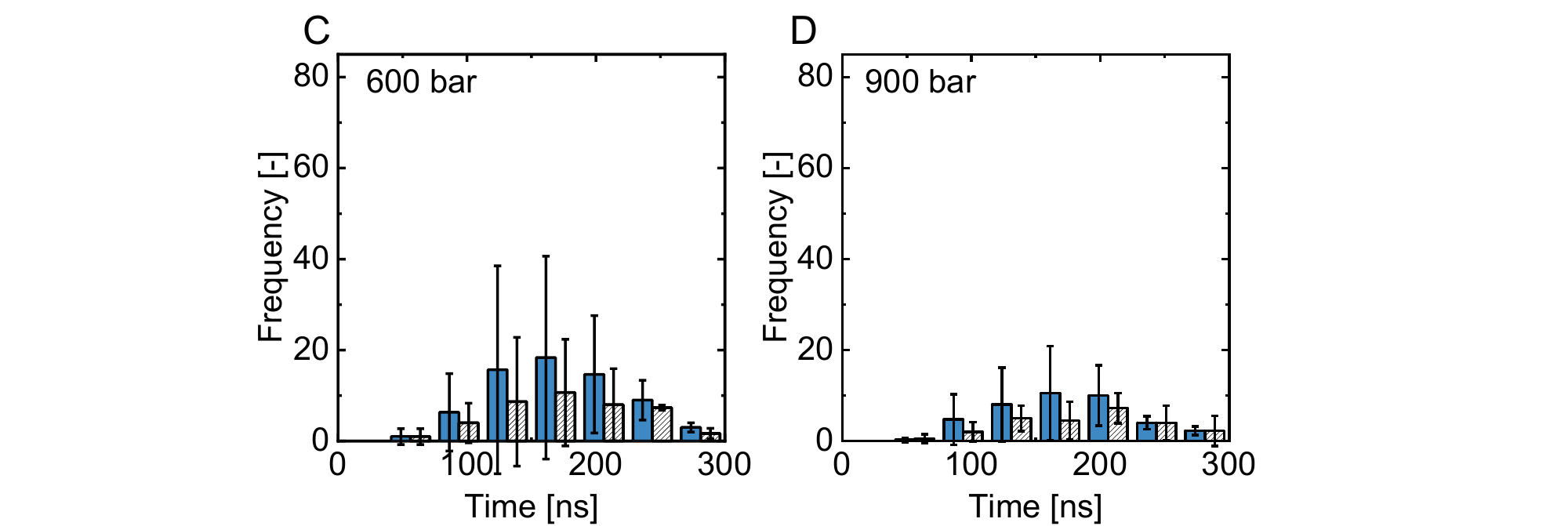}
\caption{Frequency distributions of water residence times within the membrane during transport. Blue bars indicate forward transport events, while shaded bars indicate reverse transport events. The frequency shown on the y-axis is the absolute amount of transported water molecules. Panels (\textbf{A})--(\textbf{D}) correspond to applied pressure differences of $150$, $300$, $600$, and $900\,\mathrm{bar}$, respectively. All simulations were performed with a Lattice support structure.}
\label{fig:SI_Residence}
\end{figure}

The residence times of individual water molecules inside the membrane were examined using single-water trajectories to identify complete membrane-crossing events. A water molecule was classified as forward transported if it was initially located in the feed compartment, traversed the entire membrane, and reached the permeate compartment. Conversely, a molecule was classified as reverse transported if it moved from the permeate compartment through the membrane into the feed compartment. Figure~\ref{fig:SI_Residence} shows the residence-time distributions of these completely transported water molecules, averaged over the five independently generated membrane systems. Unlike the analyses presented in the main text, all trajectory frames were considered here to maximize the number of observed complete transport events. Complete transport events were observed in both directions at all investigated pressure differences. The total number of events decreased with increasing pressure difference (Figure~\ref{fig:SI_Residence}B--D), which may be related to enhanced confinement caused by slight pressure-induced membrane compression. At the lowest pressure difference, shown in Figure~\ref{fig:SI_Residence}A, forward and reverse transport events occurred with nearly equal frequency. With increasing pressure difference, the imbalance between forward and reverse transport became more pronounced, consistent with the larger net fluxes reported in Figure~2A of the main text.

\clearpage
\section{Water cluster analysis}
\label{SI-sec:ClusterAnalysis}

\begin{figure}[ht]
\centering
\includegraphics[width=\linewidth]{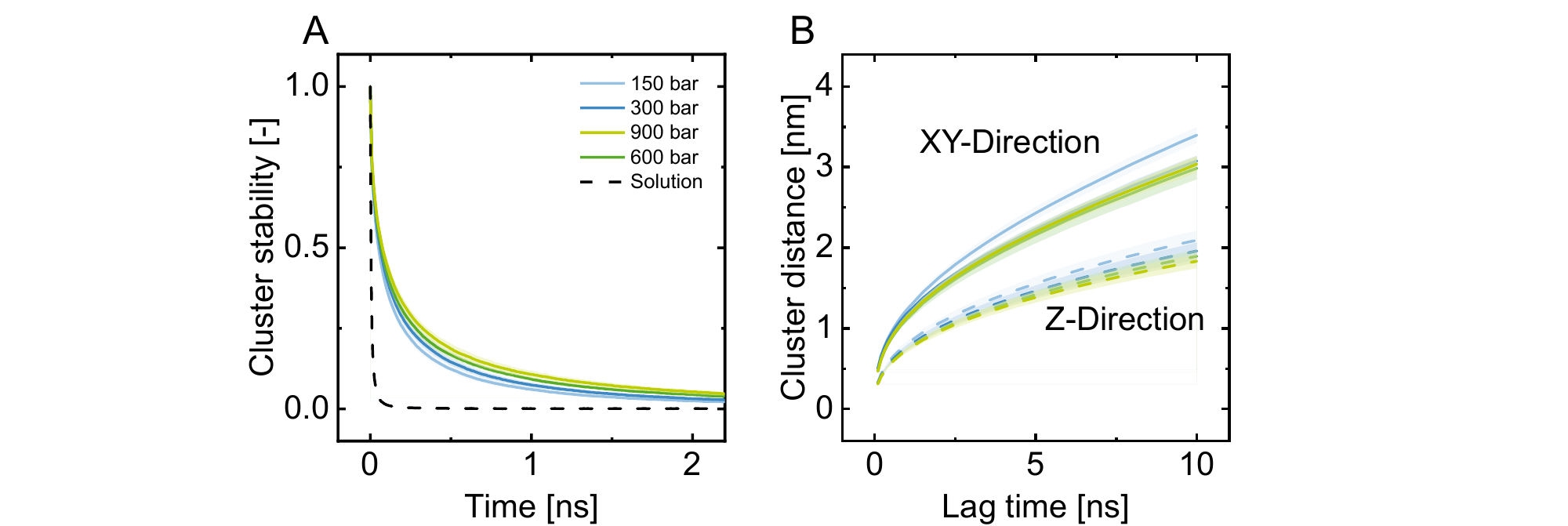}
\caption{(\textbf{A}) Survival probability of water clusters in the bulk solution and within the membrane. The survival probability quantifies the fraction of water molecules belonging to the initial neighbor shell that remain associated with the reference molecule after a given lag time, allowing for intermittent dissociation and reassociation. A value of $1$ indicates complete retention of the initial neighbor shell, whereas a value of $0$ indicates complete loss of the initially associated water molecules. (\textbf{B}) Pairwise distance between a central water molecule and the surrounding water molecules of its initial cluster as a function of lag time, resolved separately into the membrane-parallel ($xy$) and membrane-normal ($z$) directions. All results were produced with the Lattice support structure.}
\label{SI-fig:SI_ClusterSurvivial}
\end{figure}

\begin{figure}[ht]
\centering
\includegraphics[width=\linewidth]{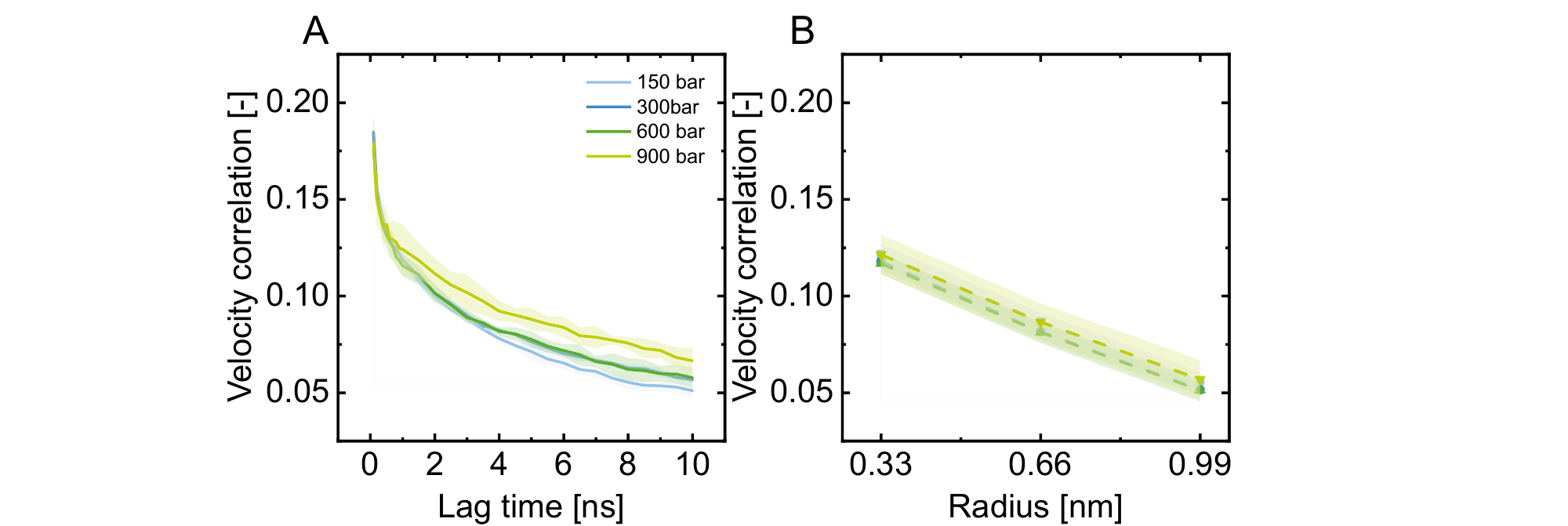}
\caption{(\textbf{A}) Directional velocity correlation between a central water molecule and water molecules initially located within its first hydration shell as a function of lag time, using a fixed coordination radius of $0.33\,\mathrm{nm}$ defined by the minimum of the oxygen--oxygen radial distribution function. (\textbf{B}) Directional velocity correlation at a fixed lag time of $1\,\mathrm{ns}$ for water molecules initially located within coordination radii of $0.33$, $0.66$, and $0.99\,\mathrm{nm}$ around the central water molecule. All results were produced with the Lattice support structure.}
\label{SI-fig:SI_ClusterVelocityCorrelation}
\end{figure}

\clearpage

\putbib[si_bib]

\end{bibunit}

\end{document}